\PassOptionsToPackage{english}{babel}

\documentclass[amsmath,amssymb,aps,twocolumn,floats,superscriptaddress]{revtex4-1}
\usepackage{hyperref}
\usepackage{amssymb,amsmath,amstext}
\usepackage{mathtools}
\usepackage{graphicx}
\usepackage{epstopdf}
\usepackage{color}
\usepackage{rotating}
\usepackage[T1]{fontenc}
\usepackage{latexsym}

\begin{document}

\title{Competition between two-photon driving, dissipation and interactions in bosonic lattice models: an exact solution}

\author{David Roberts$^{1,2}$, A. A. Clerk}
\affiliation{Pritzker School of Molecular Engineering, University of Chicago, Chicago, IL, USA \\
$^2$Department of Physics, University of Chicago, Chicago, IL, USA}

\begin{abstract} 
We present an exact solution in arbitrary dimensions for the steady states of a class of quantum driven-dissipative bosonic models, where a set of modes is subject to arbitrary two-photon driving, single-photon loss and a global Hubbard (or Kerr)-like interaction. Our solutions reveal a wealth of striking phenomena, including the emergence of dissipative phase transitions, nontrivial mode competition physics and symmetry breaking, and the stabilization of many-body $SU(1,1)$ pair coherent states. Our exact solutions enable the description of spatial correlations, and are fully valid in regimes where traditional mean-field and semiclassical approaches break down.  
\end{abstract}

\maketitle

{\it Introduction}.  
Spurred both by applications to quantum information and the advent of controllable dissipative quantum simulators 
\cite{Esslinger2010,HouckNatPhys2012,HouckPRX2017,ImamogluNatPhys2018,maDissipativelyStabilizedMott2019} there is a renewed interest in exploring driven-dissipative bosonic quantum systems in the many body limit 
(see e.g.~\cite{Diehl2008,DallaTorrePRA2013,siebererDynamicalCriticalPhenomena2013,leboiteSteadyStatePhasesTunnelingInduced2013,Hartmann2016,Biella2017,Savona2017,Dykman2018,Mora2019,rotaQuantumCriticalRegime2019}). Of particular interest are the possibility of dissipative quantum phase transitions, and the emergence of highly non-thermal steady states.  While a variety of numerical approaches have been devised to study such systems, they have limitations.   Conventional Gutzwiller mean-field approaches (see e.g.~\cite{gutz1,gutz2,gutz3,gutz4}) are unable to account for strong correlations, whereas matrix-product state methods (see e.g.~\cite{Savona2015}) are largely restricted to 1D systems.  Alternate numerical approaches for 2D exist \cite{finazziCornerSpaceRenormalizationMethod2015,Scarlatella2021}, but these can become numerically infeasible for large systems.  Given this, the ability to have exact analytic solutions for higher dimensional models would be extremely valuable.

\begin{figure}
     \centering
    \includegraphics[width=0.99\columnwidth]{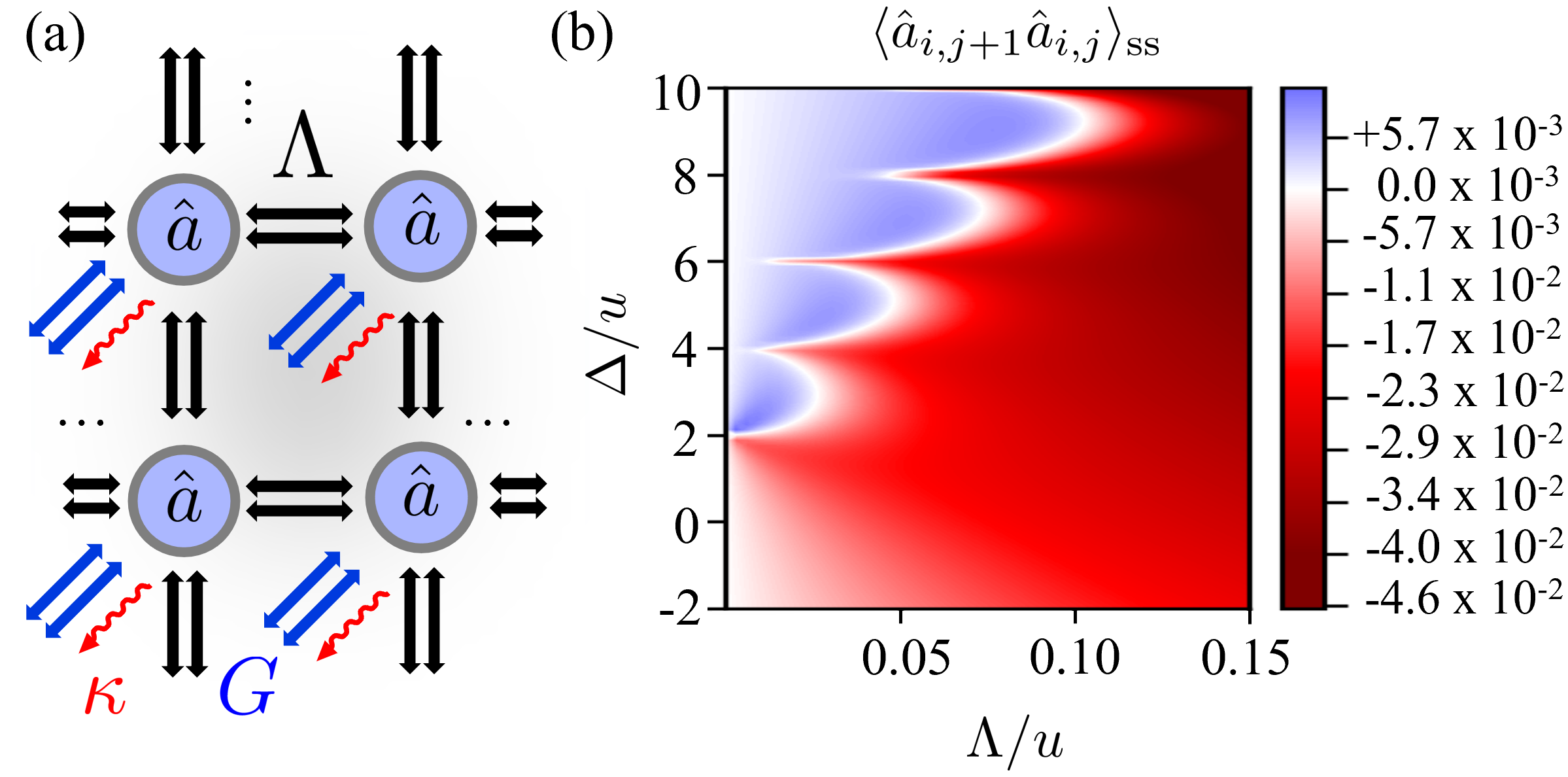}
     \caption{(a) Schematic of the model: a lattice of bosonic modes, with two-photon drives on each site ($G$) and on each nearest-neighbor (nn) bond ($\Lambda$).  There is also single-photon loss $\kappa$ on each site, and a global Hubbard (Kerr) interaction $U$.  
    (b) Our exact solution allows the description of steady-state spatial correlations.  Here, nn pairing corelations are plotted as a function of drive detuning $\Delta$ and drive amplitude $\Lambda$, for a $N=225$ site 2D lattice with $u\equiv U/N$, $\kappa = 0.01u$.  One sees clearly a Mott-lobe like structure associated with multi-photon resonances.}
    \label{fig:figone}
 \end{figure}

In this Letter, we address this outstanding challenge.  We introduce a class of strongly-interacting driven-dissipative bosonic models, and show that it is possible to {\it analytically} describe their dissipative steady states in arbitrary dimensions.  The basic system is shown in Fig.~\ref{fig:figone}:  a set of bosonic modes is subject to arbitrary two-photon driving (both on-site, and between sites), as well as to Markovian single-photon loss and a global Hubbard (Kerr) interaction that depends on total photon number.  While there are no conventional hopping interactions,
one still has a lattice structure defined by the intersite two-photon drives.  We show that the steady-state density matrix of this model is amenable to exact solution via the hidden time-reversal symmetry method \cite{robertsHiddenTimeReversalSymmetry2021,stannigelDrivendissipativePreparationEntangled2012}.  This method is related to other quantum optical solution methods \cite{drummond1980,wollinsky1988,kheruntsyan1996,kheruntsyan1997,kheruntsyan1999}, though attempts to use these in the many-body limit were unsuccessful \cite{caoTwoCoupledNonlinear2016,kheruntsyan2000}.   

Our exact solution reveals a wealth of physical phenomena. For weak driving, one sees the emergence of phase transition behaviour as system size is increased, with singularities arising in the thermodynamic limit from the merging of discrete photonic resonances. Unlike well-studied single-site models \cite{minganti2016}, the phase transition physics here can occur far from the many-photon semiclassical limit, and can show marked deviations from mean-field theory predictions.  We also show surprising connections to the representation theory of $SU(1,1)$.  Strikingly, we find that with appropriate tuning, the driven-dissipative steady state is directly related to a non-trivial many-body generalization of $SU(1,1)$ pair coherent states \cite{barutNewCoherentStates1971,luoSUCoherentStates1997,albertPaircatCodesAutonomous2019}.

We also find surprising behaviour in more strongly-driven regimes:  the system can exhibit surprising symmetry breaking phenomena and mode-competition physics, with the exact solution again providing crucial insights.  We stress that the class of models we study could be directly realized in e.g.~superconducting quantum circuits experiments, and can be viewed as a many-body extension of the driven Kerr parametric oscillator systems that are being studied extensively in the context of bosonic error correction \cite{Grimm2020,lescanneExponentialSuppressionBitflips2020}.

{\it Two-photon driven global interaction models}.
We consider a set of $N$ bosonic modes (lowering operators $\hat{a}_j$), subject to arbitrary two-photon (parametric) drives (amplitudes $M_{ij}$), as well as a global Hubbard interaction (i.e.~equal-magnitude self-Kerr and cross-Kerr interactions $U/N$).  Assuming all drives to have an identical detuning $\Delta$ from resonance, and working in the common rotating frame, the coherent system dynamics is given by:  
\begin{align}
    \hat{H}& = \frac{U}{N}\bigg(\sum_j \hat{n}_j\bigg)^2 -\Delta \sum_j \hat{n}_j+
    \sum_{i,j} \big(M_{ij} \hat{a}_i^\dagger \hat{a}_j^\dagger + h.c.\big) 
    \label{eq:Hamiltonian}
\end{align}
where $\hat{n}_j \equiv \hat{a}_j^\dagger \hat{a}_j$.   While our solution technique is more general, we focus here on the case where our modes live on the sites of a $D$-dimensional hypercubic lattice, and we have translational invariance, with $M_{ii} = G$, and off-diagonals $M_{ij} = \Lambda / 2D$ if $i,j$ are nearest neighbour sites, zero otherwise.  
This represents a modified two-photon driven Bose-Hubbard model, with single-particle hopping replaced with $p$-wave pairing terms, and the interaction made global.     
We also include dissipation:  independent Markovian single-particle loss on each site.  The full dynamics is thus described by the Lindblad master equation \begin{align}
    \partial_t \hat{\rho} = -i[\hat{H},\hat{\rho}] +\sum_j \kappa\mathcal D[\hat{a}_j]\hat{\rho}\equiv \mathcal L\hat{\rho},\label{eq:lme}
\end{align}
where $\mathcal D[\hat{X}]\hat{\rho}\equiv \hat{X}\hat{\rho} \hat{X}^\dagger-(1/2)\{\hat{X}^\dagger \hat{X}, \hat{\rho}\}$ {denotes the standard dissipative superoperator, constructed from an arbitrary linear operator $\hat{X}$ acting on the Hilbert space of our system}. We note that related two-photon driven many-body bosonic models have been recently studied numerically \cite{Savona2017,Mora2019,rotaQuantumCriticalRegime2019}.  

{Eq.~\ref{eq:Hamiltonian} exhibits a generic tension common to many driven-dissipative systems.  The drives favour populating the system with pairs of photons, creating squeezing correlations.  This is opposed by the losses, the energy detuning $\Delta$ (which makes pair addition non-resonant), and most crucially the interaction $U$ (which is like a number-dependent detuning).  This yields the possibility of phase transitions, where a high density could self-consistently make the drives resonant.  While there is no conventional hopping, the nonlocal pair drives can create spatial correlations (and are like an "Andreev-reflection" hopping process).  Note that our model could be realized in a variety of setups including superconducting circuits and more conventional quantum optical platforms (see \cite{supp} for a simple circuit implementation of our model).  We also note that our solution is even more general than Eq.~\ref{eq:Hamiltonian}.  As shown in \cite{supp}, for a given set of drive amplitudes $M_{ij}$, there exist a class of standard hopping terms that can be added to $\hat{H}$ without changing the dissipative steady state.  We can thus describe, e.g., bipartite lattices with local hopping and pairing terms.  }

Our goal in this work is to understand the dissipative steady state 
$\hat{\rho}_\text{ss}$ of our system, which satisfies $\mathcal L\hat{\rho}_\text{ss} = 0$.  Surprisingly, for all parameter values and dimensionalities, this can be done exactly and analytically, using the hidden TRS (hTRS) / coherent quantum absorber approach introduced in \cite{stannigelDrivendissipativePreparationEntangled2012,robertsHiddenTimeReversalSymmetry2021}.  This method postulates the existence of an anti-unitary operator $\hat{T}$,  in terms of which the associated purification of $\hat{\rho}_\text{ss}$ (which lives in a doubled Hilbert space)
\begin{align}
    \hat{\rho}_\text{ss} \equiv \textrm{Tr}_R 
    |\Psi_{\hat{T}}\rangle \langle\Psi_{\hat{T}}|,
    \,\,\,\,\,\,\,\,\,
   |\Psi_{\hat{T}}\rangle \equiv \sum_n \sqrt{p_n} |n\rangle_L \hat{T}|n\rangle_R,\label{eq:tracingout}
\end{align}
satisfies a generalized symmetry constraint \cite{robertsHiddenTimeReversalSymmetry2021}. Here $|n\rangle,p_n$ are the eigenvectors and eigenvalues of $\hat{\rho}_\text{ss}$, $L$ denotes states in the physical Hilbert space, and $R$ denotes states in the auxiliary Hilbert space, {which is another copy of the physical Hilbert space}. The ansatz that $\hat{T}$ is a hTRS implies a set of conditions on $|\Psi_{\hat{T}}\rangle$ that must be solved.  For this system, this can be done analytically \cite{supp}.

The resulting solution for the pure state 
$|\Psi_{\hat{T}} \rangle$ has a striking form.  It describes an unusual kind of pair condensate:  all particles occupy the same two-body wavefunction whose spatial structure is determined by the driving amplitudes $M_{ij}$. We find \cite{supp}:
\begin{align}
    &|\Psi_{\hat{T}}\rangle =\sum_{m=0}^\infty 
    \frac{c_m}{m!} 
    \left( \hat{K}_+ \right)^m |\Omega\rangle,
    \,\,\,\,\,\,\,
    \hat{K}_+ :=\frac{N}{2U}\sum_{ij}M_{ij}\hat{\alpha}_i^\dagger\hat{\alpha}_j^\dagger,
    \label{eq:TFDsoln}
\end{align}
where $\hat{K}_+$ is the effective pair creation operator, $\hat{\alpha}_j\equiv (\hat{a}_{j,L} +\hat{a}_{j,R})/\sqrt{2}$, and $|\Omega\rangle\equiv |0\rangle_L|0\rangle_R$ is the vacuum. The coefficients $c_m$ in the expansion take the simple form
\begin{align}
    c_m \propto (-1)^m/(\delta)_m,
\end{align}
where $(\delta)_m := \delta(\delta+1)\cdots (\delta+m-1)$ denotes the Pochhammer symbol (rising factorial), and where the dimensionless detuning parameter $r$ is
\begin{align}
    \delta:=1-N\Delta_\text{eff}/2U, \,\,\,
    \Delta_\text{eff} := \Delta + i\kappa/2.
    \label{eq:rDefinition}
\end{align} 
We stress that when $U \neq 0$, this pure-state pair condensate is  highly non-Gaussian and exhibits Wigner-function negativity. The parameter dependence of this state is also remarkable. The global Hubbard interaction $U$ along with the detuning $\Delta$ and loss $\kappa$ determine the effective "fugacity" of our pair gas via the $c_m$ coefficients.  In contrast, all spatial structure (encoded in $M_{ij}$) is encoded completely in the two-body ``wavefunction'' of each paired boson. {Finally, the resulting dissipative steady state is non-thermal, in that it cannot be written as $\exp(-\beta \hat{H})$ for some $\beta$ \cite{supp}}.

\begin{figure}
     \centering
    \includegraphics[width=0.99\columnwidth]{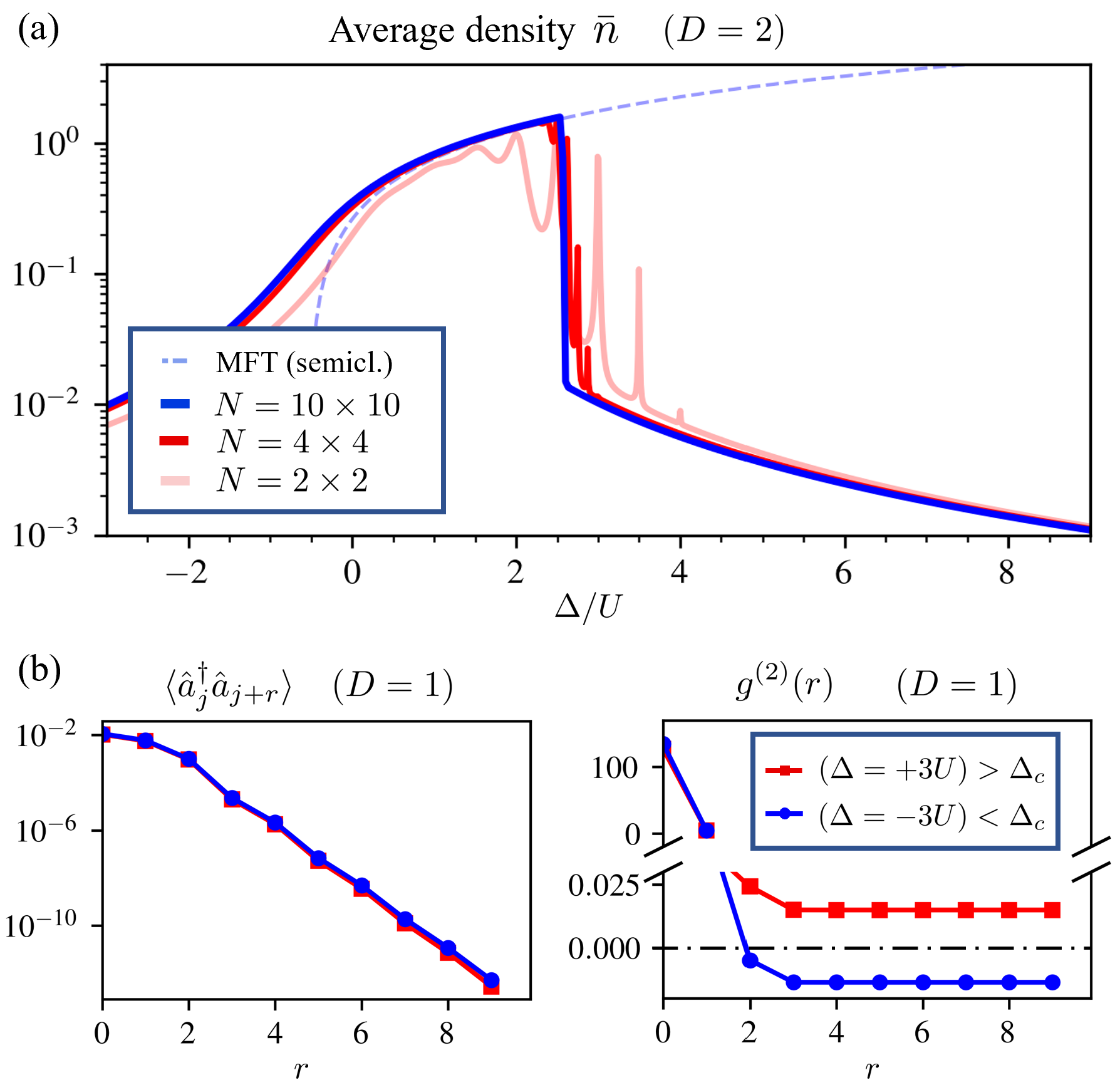}
     \caption{\textbf{Driven-dissipative phase transitions.}
(a) Average density $\bar{n}$ versus detuning $\Delta$ for various sized 2D square lattices (periodic boundary conditions, $\kappa = 0.01U$,  $G = U/5, \Lambda = U/4$). As system size increases, discrete resonances merge to yield a jump in the density and a first-order phase transition. {We also show the predictions of a basic semiclassical mean-field theory, which predicts a zero-density solution that cannot be shown here due to the log scale on the $y$-axis.}
(b) Here, we attempt to distinguish the bunched (red squares) and antibunched (blue circles) phases via their correlations, respectively single-particle (left panel) and density-density (right panel) correlations. We choose $\Delta = +3U$ as representative of the bunched phase and $\Delta = -3U$ as representative of the antibunched phase. Both plots show data for a $N = 100$ site periodic lattice with $D =1$. All other parameters are the same as in panel (a).  All results are computed using the exact solution in Eq.~\eqref{eq:TFDsoln}.}
     \label{fig:fig_four_reborn}
 \end{figure}

{\it Emergence of phase transitions}. 
The exact solution allows us to study the emergence of dissipative phase transitions as the number of sites $N$ becomes large, i.e.~in the thermodynamic limit.  This can be done for arbitrary dimensionality $D$, and while still remaining in low-density regimes where semiclassical approximations would fail.  
We find a direct connection between first-order phase transitions that occur at large $N$, and discrete multi-photon resonances that can be resolved at smaller $N$.  This is seen clearly in Fig.~\ref{fig:fig_four_reborn}(a), which shows the average steady-state photon density versus $\Delta$ in a $D=2$ model, for different system sizes. The discrete resonances at modest $N$ occur when the dimensionless detuning $r$ is close to a negative integer.  The exact solution tells us that when $\delta=-n +\epsilon$ with $|\epsilon| \ll 1$, the relative "fugacity" between the $n+1$ and $n$ pair configurations diverges as $\epsilon \rightarrow 0$:  
$c_{n+1} / c_n =-1 / (n+\delta) = O(\epsilon^{-1})$. This divergence (cut-off by $\kappa$) leads to an enhanced photon number, and thus sharply-defined resonances occuring at detunings
$\Delta_n = 2U (n+1) /N$
(see Fig.~\ref{fig:fig_four_reborn}(a)). As $N\to \infty$, the spacing between resonances vanishes, leading to a first-order phase transition where the density exhibits a jump as a function of $\Delta$. 
Fig.~\ref{fig:fig_four_reborn}(a) also shows a comparison against the predictions of a simple semiclassical mean-field theory (see \cite{supp} for more details, as well as comparisons to Gutzwiller mean field theory).  

A further virtue of the exact solution is that it gives full access to spatial correlations.  We find that these correlations provide a much better way of distinguishing phases compared to purely local observables.  In the large-$N$ limit, two-point equal-time correlators in the steady state such as $\langle \hat{a}_{i+r}\hat{a}_i\rangle_\text{ss},\langle \hat{a}_{i+r}^\dagger\hat{a}_i\rangle_\text{ss}$ always decay exponentially with distance (see Fig.~\ref{fig:fig_four_reborn}(b)).  
In stark contrast, the global Hubbard interaction generates long-range (but weak) density-density correlations.  To study this quantitatively, we define in $D=1$ the reduced density-density corelator 
\begin{align}
    g^{(2)}(i,r) := \frac{\langle \hat{n}_{i+r}\hat{n}_{i}\rangle_\text{ss}-\bar{n}^2}{\bar{n}^2}.
    \label{eq:g2Definition}
\end{align}
Here, $\bar{n}\equiv \langle \hat{n}_j\rangle_\text{ss}$ is the mean onsite occupation in the steady state, and we note that
$g^{(2)}(i,r)$ is independent of $i$ away from boundaries.  An analogous definition holds for $D>1$.

\begin{figure}
     \centering
    \includegraphics[width=0.99\columnwidth]{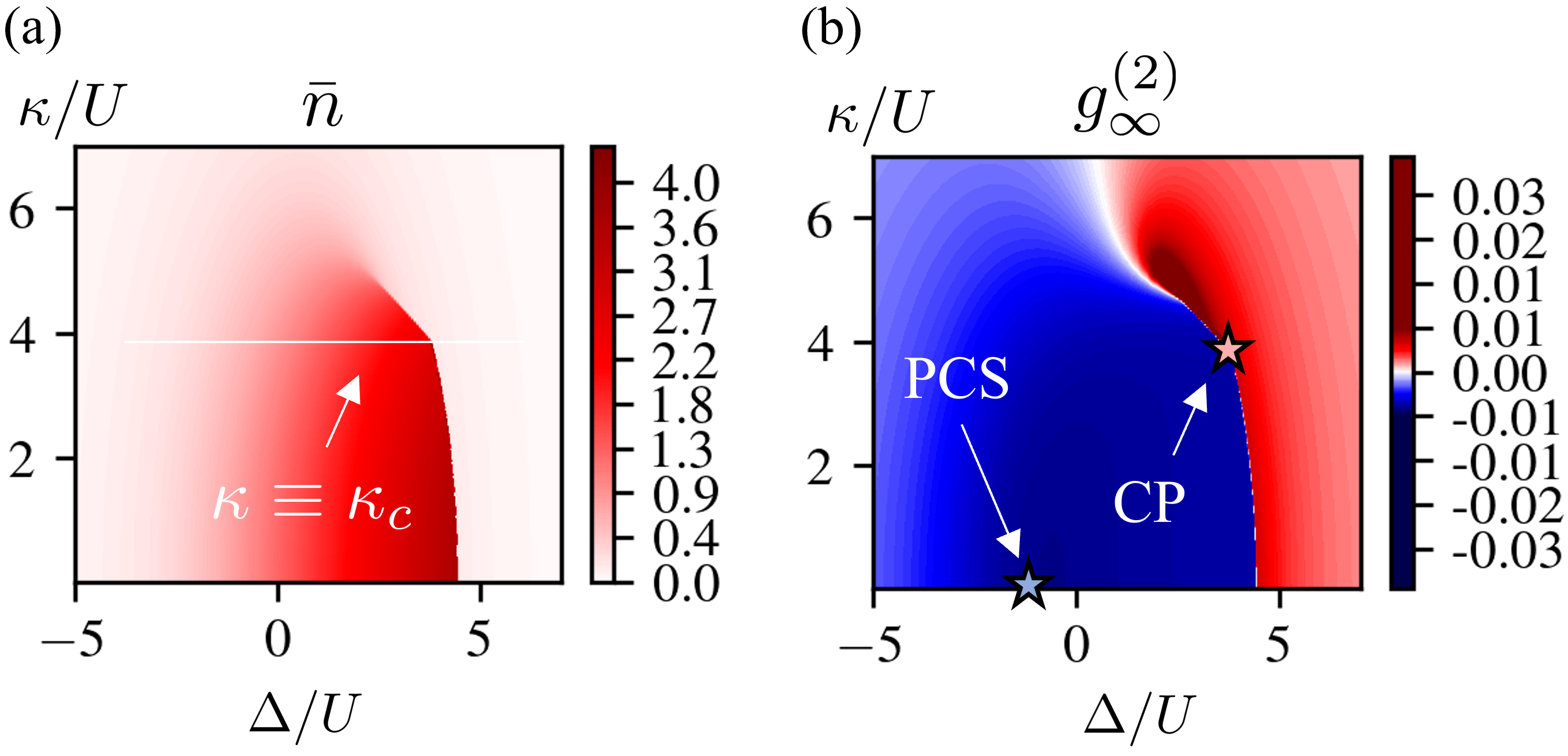}
     \caption{\textbf{Phase diagram for $D = 0$}. (a) Average density as a function of detuning $\Delta$ and loss $\kappa$, with $N=500$, $\Lambda = 0$, and $G = U$.  Phase boundaries can be seen, the critical damping value $\kappa_c$ is also indicated: for $\kappa > \kappa_c$, the first order PT vanishes.  
     (b) Asymptotic long-distance behavior of the density-density correlation function, as captured by $g^{(2)}_\infty$
     (c.f.~Eq.~\eqref{eq:g2Definition}); the sign of this quantity more clearly distinguishes the two relevant phases in the model. A critical point $\Delta_\text{eff}^c := \Delta_c +i\kappa_c/2$ marks the exact location where $g^{(2)}_\infty$ becomes continuous across the phase boundary. Same parameters as in panel (a).  The parameter tuning that results in a many-body pair coherent state is indicated with a star.} 
     \label{fig:integrability}
 \end{figure}

We find that the two phases of our model can be cleanly distinguished by the sign of the large-distance density-density correlations, i.e. by
$    g^{(2)}_\infty \equiv \lim_{|r|\to \infty}g^{(2)}(r).$ We call the phase where $g^{(2)}_\infty  > 0$ a "bunched" phase, where density fluctuations are positively-correlated at long distances, and the remaining phase with $g^{(2)}_\infty < 0$ an "antibunched" phase. When $\kappa$ is sufficiently small, these phases are connected by the first-order phase transition mentioned above.  The corresponding jump in density is accompanied by a sign change in $g^{(2)}_\infty$, see Fig.~\ref{fig:fig_four_reborn}(b), right panel.
We also note that for modest values of $N$, the multiphoton resonance physics described above can also lead to interesting structures resembling Mott lobes \cite{leboiteSteadyStatePhasesTunnelingInduced2013,leboiteBoseHubbardModelRelation2014}, if one looks at intersite correlations.  This is shown in Fig.~\ref{fig:figone}(b).

{\it Criticality in the $D=0$ model}. 
The above physics becomes especially clear in the limit where $\Lambda \equiv 0$, i.e. purely local driving.  There is no remaining spatial structure, hence we call this the $D=0$ limit.  As we saw in Fig.~\ref{fig:fig_four_reborn}, for $D>0$, our model has a finite correlation length characterizing the decay of two-point correlators.  The $D=0$ model sets this length to zero, while retaining the more interesting physics associated with density-density correlations. 
The $D=0$ limit is also experimentally relevant:  it can be realized directly using a relatively simple superconducting circuit \cite{supp}.

The $D=0$ case has another key virtue: it allows a dramatic simplification in the calculation of observables,  as now $\hat{K}_+,\hat{K}_+^\dagger$, and $\hat{K}_z\equiv (2NG^2/U)[\hat{K}_+^\dagger,\hat{K}_+]$ form a representation of the Lie algebra of $SU(1,1)$. This makes the problem of evaluating moments with respect to the state $|\Psi_{\hat{T}}\rangle$ given in Eq.~\eqref{eq:TFDsoln} completely algebraic;  one only requires knowledge of the bosonic representation theory of $SU(1,1)$.  Further, harmonic analysis in $\mathbb{R}^N$ yields a satisfactory characterization of the requisite representation theory \cite{supp}.  We are thus able to compute local observables and correlators for systems with tens of thousands of sites and at unit density. For our $D=0$ model and for large $N$, we can verify by brute force that $\lim_{\Delta \to \Delta_c^\pm} \text{sign}\{g^{(2)}_\infty\} = \pm 1$, where $\Delta_c$ denotes the location of the discontinuity in $\bar{n}$.  This confirms that the first-order PT marks the boundary between bunched- and antibunched phases (c.f. Figure \ref{fig:fig_four_reborn}). 
We also find that this first-order PT only exists when $\kappa < \kappa_c$, where $\kappa_c$ is a critical damping threshold, {akin to a critical pressure in a liquid-gas transition (c.f. Figure \ref{fig:integrability}a). As in a liquid-gas transition, above the critical point the two phases are smoothly connected, as is indicated by the continuity of $g^{(2)}_\infty$ in Figure \ref{fig:integrability}b. Here, we use the exact solution to estimate $\kappa_c$, by explicitly observing the divergence of the susceptibility $\chi \equiv \partial \bar{n}/\partial \Delta$ as $\kappa\to\kappa_c^+$; see \cite{supp} for more details. }

 
{\it Many-body pair-coherent states}. When $D>0$, analysis based on the exact solution becomes more challenging. One can still obtain a representation of the Lie algebra $SU(1,1)$ by defining a (generalized) pair-lowering operator
$    \hat{K}_- :=\frac{U}{2N}\sum_{ij}(M^{-1})_{ij}\hat{\alpha}_i\hat{\alpha}_j$, which has the effect of removing a pair of bosons: $\hat{K}_-\hat{K}_+^m|\Omega\rangle \propto \hat{K}_+^{m-1}|\Omega\rangle$. However, $\hat{K}_-$ is not equal or proportional to $\hat{K}_+^\dagger$ unless $D\equiv 0$ or $D\equiv \infty$.
The result is that representation-theoretic techniques are of no utility when $D > 0$.  Nonetheless, the Lie-theoretic point of view is still useful in helping reveal unusual phenomena.  

In particular, at special detuning values, the gas of boson pairs constituting the purification of the steady state (c.f.~Eq.~\eqref{eq:TFDsoln}) forms a many-body pair coherent state (PCS), that is, an eigenstate of the operator $\hat{K}_-$ \cite{barutNewCoherentStates1971}. From the form of the solution, we see that this happens when $c_{m+1}/c_m =-k/(N/2 + m)$, where $k=-1$ is the corresponding eigenvalue of $\hat{K}_{-}$ \cite{supp}.
From Eq.~\eqref{eq:rDefinition}, we see that this requires $\delta = N/2$, corresponding to $\kappa \rightarrow 0^+$ and 
\begin{align}
    \Delta \rightarrow \Delta_{PCS} := U(2-N)/N
\end{align}
Note that for the case of just a single mode $N=1$, this corresponds to the known physics of a Kerr parametric oscillator \cite{wollinsky1988}.  In this case, $\Delta = U$ is the same as zero detuning if one normal-orders the Kerr interaction, and $|\Psi_{\hat{T}}\rangle$ reduces to an even-parity cat state. 

We stress that there are observable consequences associated with the formation of this many-body PCS.  As one approaches the special detuning, there are no fluctuations in the global pairing, as quantified by the operator $\hat{K}_{-}$.  One can explicitly show that:
\begin{align}
    &\Bigg\langle \bigg(\sum_{ij} (M^{-1})_{ij}\hat{a}_i\hat{a}_j\bigg)^{\dagger n}\bigg(\sum_{ij} (M^{-1})_{ij}\hat{a}_i\hat{a}_j\bigg)^{m}\Bigg\rangle_\text{ss}\nonumber\\
    &~~~~~~~~~~~~~~~~~~~~~~~~\underset{\Delta\to \Delta_{PCS}}{\propto} k^{*n}k^m.
\end{align}
Similar to their two-mode counterparts \cite{agarwal1988,agarwal2005}, the many-body PCS we describe here may have utility for bosonic quantum error correction \cite{mirrahimiDynamicallyProtectedCatqubits2014,leghtasConfiningStateLight2015,puriStabilizedCatDriven2019,lescanneExponentialSuppressionBitflips2020,Grimm2020}. We note that the many-body PCS that emerge here are distinct from the multi-mode states discussed in Ref.~\cite{albertPaircatCodesAutonomous2019}.

\begin{figure}
     \centering
    \includegraphics[width=0.99\columnwidth]{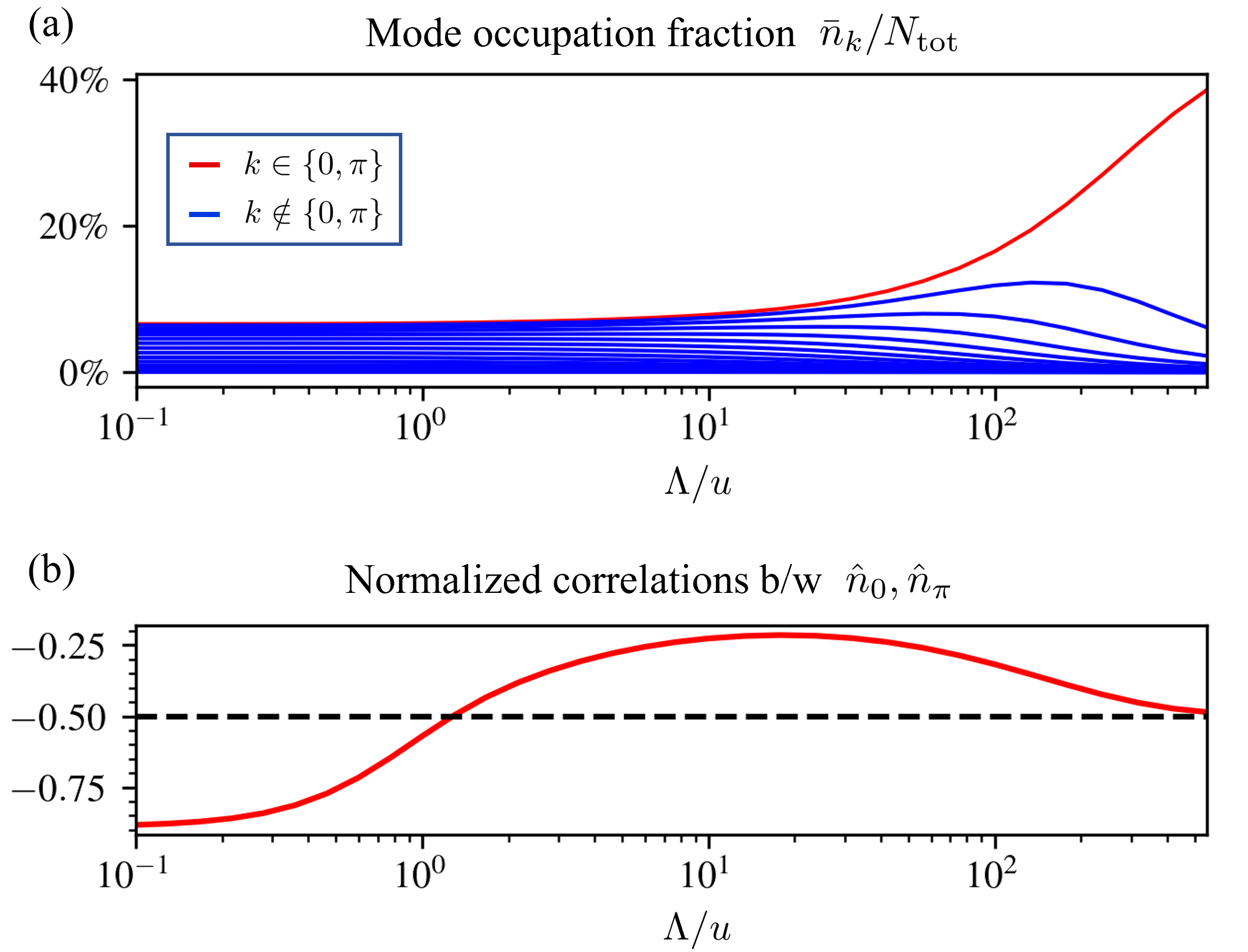}
     \caption{\textbf{Symmetry breaking at strong driving}.
     (a)  Occupancy $\bar{n}_k$ of standing wave modes in a odd-length $D=1$ open chain, as the drive $\Lambda$ is increased.  For large drives, the modes with the largest pairing amplitudes, $k=0,\pi$, dominate. $N_\text{tot}$ denotes average total photon number. 
     Parameters are $\Delta = 0,\kappa = u/100,$ $u \equiv U/N$, $N = 31$.  
     (b) Normalized density correlations between the modes at $k=0,\pi$ (red curve), and the horizontal asymptote $y \equiv -1/s$ predicted by a uniform sphere distribution (black dashed line). Here, $s=2$. Parameters same as in panel (a).}
     \label{fig:sc_fig}
 \end{figure} 

{\it Symmetry breaking.} In the strong-driving regime, our model exhibits a surprising symmetry breaking phenomenon.  First, note that the singular values of our matrix of pair-driving amplitudes is $\lambda_\mathbf{k} = \frac{1}{u}\bigg|\frac{\Lambda}{D}\sum_{j = 1}^D \cos k_j + G\bigg|$ where the wavevector $\mathbf{k}$ labels standing wave modes.  Let $\lambda_*$ denote the maximum singular value, and $s$ denote the number of distinct modes that it corresponds to (so-called max pairing modes).  For large driving, one can analytically show that the steady state Wigner function 
$W[\{\alpha_\mathbf{k} \}]$
corresponds to a uniform distribution over the $(s-1)$-sphere defined by 
\begin{align}
    \sum_{\lambda_\mathbf{k} = \lambda_*}x_{\mathbf{k}}^2= \text{const}.,~~~~~x_\mathbf{k}\equiv e^{-i\theta}\alpha_\mathbf{k}  
\end{align}
with $x_\mathbf{k}\in \mathbb{R}$ and $\theta$ an overall phase \cite{supp}. Even though there is a near continuum of pairing eigenvalues, for large driving, the max pairing modes completely dominate.  This behaviour is shown explicitly in Fig. \ref{fig:sc_fig}(a).  The structure of this solution also directly leads to an anti-correlation between mode amplitudes that is purely geometric, see Fig. \ref{fig:sc_fig}(b).

  The mode selection in our system can be related to spontaneous symmetry breaking. Real rotations amongst the max-pairing modes form a non-abelian group of weak symmetries isomorphic to $O(s,\mathbb{R})$ which commutes with the Lindbladian $\mathcal L$.  At high driving strengths we conjecture that this symmetry is spontaneously broken. This is seen clearly at the semiclassical level, where one can show \cite{supp} that every point on the max pairing $(s-1)$-sphere is a {\it stable} stationary state of the dynamics.  Each such solution of course breaks the underlying mode-rotation symmetry. In the full quantum theory, fluctuations lead to a slow randomization on this space of symmetry broken solutions, yielding the final unique steady state.    
  The effective mode selection phenomena  in our model is reminscient (but not identical) to analogous effects in other systems (see e.g.
\cite{narducciModemodeCompetitionUnstable1986,gongNumericalModelingTransverse2007,stoneConversionEfficiencyKerrMicroresonator2022}). {Ref.~\cite{carlos_PRL_2021} also describes (using semiclassical MFT) related phenomena in a many-mode model with uniform pairing, with mode selection being controlled by dispersion as opposed to pairing amplitudes.  We stress that in contrast to \cite{carlos_PRL_2021} our exact solution lets us describe all quantum fluctuation effects, allowing analytical insights into how our mode selection effect emerges as the dimensionless driving rates $G/U,\Lambda/U$ become large, see Fig. \ref{fig:sc_fig}(a)}.

{\it Discussion}. 
We have introduced a class of strongly interacting, two-photon driven bosonic lattice models whose dissipative steady states can be found exactly.  The models exhibit a wealth of interesting phenomena, including emergent phase transitions, many-body pair coherent states, and novel mode competition and symmetry breaking.  Our work provides an important means for benchmarking approximation techniques, and also reveals that the physics of Kerr parametric oscillators (studied extensively for error correction) is even richer in the many body limit.  It also suggests that the hTRS solution method could be used to successfully address a host of truly many-body problems.

We thank Alexander McDonald, Qian Xu and Mark Dykman for helpful discussions.  This work was supported by the Air Force Office of Scientific Research MURI program under Grant No.~FA9550-19-1-0399, and the Simons Foundation through a Simons Investigator award (Grant No. 669487).   

\bibliographystyle{apsrev4-1}
\bibliography{ms}

\end{document}


\selectlanguage{english}

\title{Supplemental Material for\\ "Competition between two-photon driving, dissipation and interactions in bosonic lattice models: an exact solution"}

\author{David Roberts$^{1,2}$, A. A. Clerk}
\affiliation{Pritzker School of Molecular Engineering, University of Chicago, Chicago, IL, USA \\
$^2$Department of Physics, University of Chicago, Chicago, IL, USA}
\date{\today}

\maketitle

\onecolumngrid

\tableofcontents

\section{Augmenting the model with hopping terms}

\begin{figure}
    \centering
    \includegraphics[width = 0.9\columnwidth]{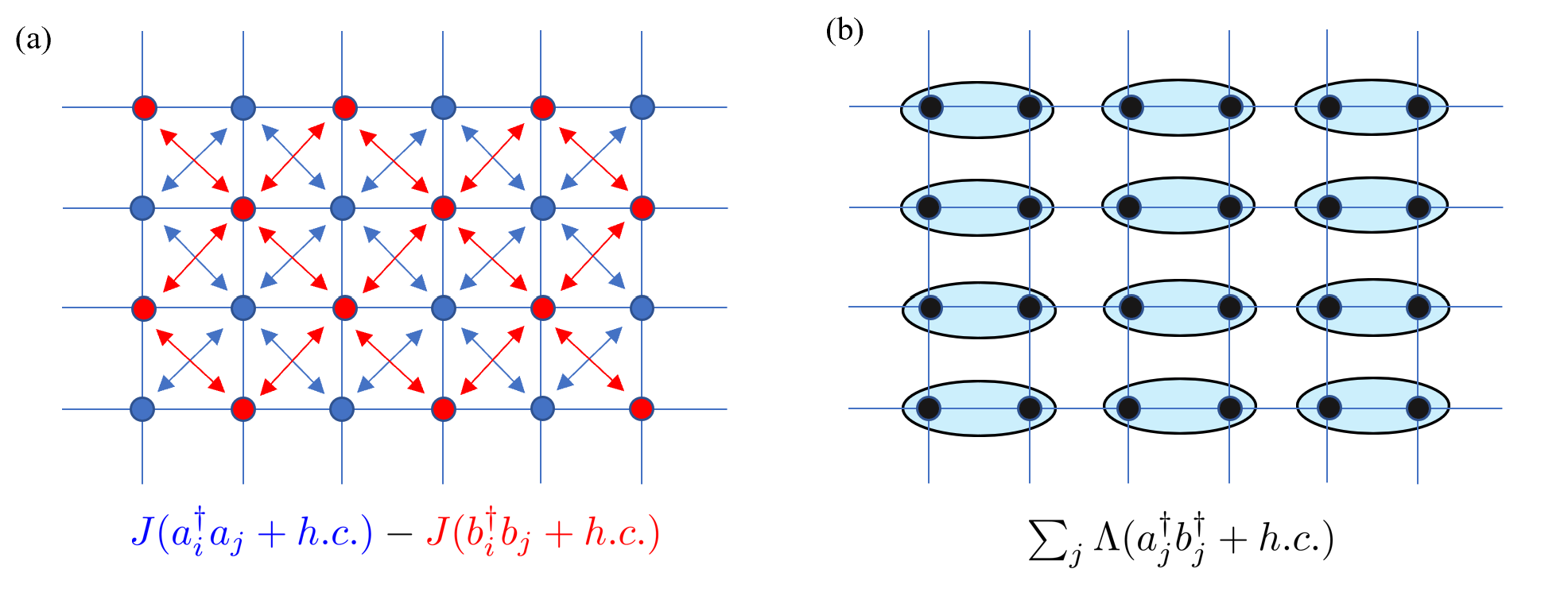}
    \caption{{\bf Exactly-solvable Bose-Hubbard models with hopping and pair-driving}. (a) A bipartite lattice is subjected to chirally-symmetric hopping (depicted using blue and red arrows) as well as pair-driving of the form $\delta \hat{H} = \sum_j(\hat{a}_j^\dagger\hat{b}_j^\dagger + h.c.)$ (not shown in this subpanel), and an infinite-range Bose-Hubbard interaction (not shown in this subpanel). (b) At late times, the steady state becomes stationary with respect to the hopping process depicted in panel  (a), and so the steady state can be analyzed by ignoring the hopping processes and considering only the pair-driving process (depicted by blue ovals) and the Hubbard interaction (not shown in this subpanel) in isolation.}
    \label{fig:fig6}
\end{figure}

{ The driven-dissipative bosonic lattice model that we solve in this work is given by the following Lindblad master equation:
\begin{align}
    \partial_t \hat{\rho}& = -i[\hat{H}, \hat{\rho}]+\sum_{j} \mathcal D[\hat{L}_j],~~~~~\hat{H} = u\hat{N}^2 -\Delta \hat{N} + \sum_{ij} \big(M_{ij} \hat{a}_i^{\dagger }\hat{a}_j^\dagger + h.c.\big),~~~~~~~
    \hat{L}_j = \sqrt{\kappa}\hat{a}_j\label{eq:equation_one}
\end{align}
where here, $\hat{N}\equiv \sum_{j=1}^N \hat{a}_j^\dagger \hat{a}_j$ denotes total boson number, and $u\equiv U/N$. In addition, we have an arbitrary complex-valued two-mode squeezing
array $M_{ij}$. 

We will now describe how single-photon tunneling terms of the form $J\hat{a}_i^\dagger \hat{a}_j + h.c.$ can be added to the Hamiltonian while preserving its solvability. A proper explanation of this necessitates a discussion of the symmetries of the dissipative evolution generated by the Lindbladian $\mathcal L$. We begin by discussing the two-mode squeezing array $M$. We define the {\it symmetry group} $S$ of the array $M$ to be the group formed by unitary beam-splitter transformations that leave the pairing array invariant. Under such a transformation, represented by a unitary matrix $W$, the pairing array $M$ transforms as $M\to WMW^T$. Therefore 
\begin{align}
    S &= \bigg\{ W\in U(N):~WMW^T = M\bigg\}=V\prod_{\lambda \in \Sigma} O(s_\lambda,\mathbb{R})\,V^\dagger,\label{eq:result}
\end{align}
where $M=V\Sigma V^T$ is the Autonne-Takagi factorization of the matrix $M$ with $\Sigma$ the diagonal matrix of singular values, and where $s_\lambda$ denotes the degeneracy of the singular value $\lambda\in \Sigma$. In arriving at the result \eqref{eq:result} we have used the fact that the symmetry group of $\Sigma$ is the product group $\prod_{\lambda \in \Sigma} O(s_\lambda,\mathbb{R})$. Since $S$ by definition conserves total particle number, any beam-splitter transformation that lies in $S$ is a weak symmetry of the Lindbladian \eqref{eq:equation_one}.

Let us now assume that $\mathcal L$ has a unique steady state. Then this steady state must respect the weak symmetries of the Lindbladian $\mathcal L$ \cite{bua2012}. In particular, the steady state is invariant under the symmetry group $S$ of the pairing array. In particular, we can add any element in the Lie algebra $\mathfrak{s}$ of $S$ to the Hamiltonian in \eqref{eq:equation_one} while still preserving the steady state. The Lie algebra of $S$ can be explicitly computed:
\begin{align}
    \mathfrak{s} = \mathcal V\bigg\{i\sum_{\lambda \in \Sigma}\sum_{i,j \in \underline{s}_\lambda} t_{ij}(\hat{a}_i^\dagger \hat{a}_j-\hat{a}_j^\dagger \hat{a}_i),~~~t_{ij}\in \mathbb{R} \bigg\}\mathcal V^\dagger.
\end{align}
where $\underline{s}_\lambda$ denotes the subset of mode indices corresponding to the singular value $s_\lambda$, and $\mathcal V$ is some beam-splitter transformation that implements the unitary transformation $V$, i.e. $\mathcal V^\dagger\hat{a}_i\mathcal V\equiv\sum_j V_{ij}\hat{a}_j$.

\subsection{Imaginary hopping arrangements}
When the singular values of the pairing array are fully degenerate, the pairing array is maximally symmetric: in this case, the symmetry group $S$ is, up to the unitary similarity transform $V$, just the full rotation group as a subgroup of the unitary group. The simplest example is the case where the pairing array is diagonal and uniform: $M_{ij} = G\delta_{ij}$, which lets us set $V = 1$. Therefore, in this case, the symmetry group of the pairing array is generated by terms of the form
\begin{align}
    \delta\hat{H} \equiv i\sum_{i,j=1}^N t_{ij}(\hat{a}_i^\dagger \hat{a}_j-\hat{a}_j^\dagger \hat{a}_i).\label{eq:equation_four}
\end{align}
Recall that, since $\delta\hat{H}$ generates a weak symmetry of $\mathcal L$, one can add $\delta\hat{H}$ to the Hamiltonian without changing the steady state of $\mathcal L$. That is, one can add any set of imaginary hopping terms to the Hamiltonian without changing the steady state of $\mathcal L$.

\subsection{Chirally-symmetric hopping arrangements}
Thanks to the similarity transform $V$ appearing in the symmetry group $S$, there are many situations where our solvable model can be augmented with conventional hopping terms.  As a particularly striking example of this, consider a global Bose-Hubbard model defined on a bipartite lattice with $2N$ sites, with corresponding modes $\hat{a}_1,\dots \hat{a}_N,\hat{b}_1,\dots, \hat{b}_N$. We can consider a special case of the Lindbladian \eqref{eq:equation_one} where the pairing array is fully "dimerized", that is, where the Hamiltonian takes the form
\begin{align}
    \hat{H}=u\hat{N}^2-\Delta \hat{N} + \Lambda\sum_{j} (\hat{a}_j^\dagger \hat{b}_j^\dagger + h.c.).\label{eq:dimerized}
\end{align}
In this case, one can augment the above Hamiltonian with the following hopping terms:
\begin{align}
    \delta\hat{H}\equiv \sum_{ij}J_{ij}\bigg((\hat{a}_i^\dagger \hat{a}_j + h.c.) -(\hat{b}_i^\dagger \hat{b}_j + h.c.)\bigg).\label{eq:equation_three}
\end{align}
Note that such a model describes arbitrary hopping within the $A$-sublattice, which is mirrored in the $B$-sublattice. This situation is schematically depicted in Fig. \ref{fig:fig6}a. To see explicitly that the hopping terms generate rotational symmetries of $\mathcal L$, we have to diagonalize the pairing array. In particular, one can do this by defining modes 
\begin{align}
    \hat{c}_{j,+}^\dagger\equiv \frac{\hat{a}_j^\dagger+\hat{b}_j^\dagger}{\sqrt{2}},~~~\hat{c}_{j,-}^\dagger\equiv \frac{\hat{a}_j^\dagger-\hat{b}_j^\dagger}{i\sqrt{2}}.
\end{align}
In this mode basis, the Hamiltonian, with the hopping terms included, reads as
\begin{align}
    \hat{H}=u\hat{N}^2-\Delta \hat{N} + \Lambda\sum_{j, \pm}(\hat{c}_{j,\pm}^{\dagger 2} + h.c.) + i\sum_{i,j=1}^N J_{ij}(c_{i,-}^\dagger \hat{c}_{j,+} - h.c.) +  i\sum_{i,j=1}^N J_{ij}(c_{i,+}^\dagger \hat{c}_{j,-} - h.c.).
\end{align}
In this diagonal mode basis, it is extremely clear that the hopping terms generate rotations of the modes $\hat{c}_{j,\pm},\hat{c}_{j,\pm}^\dagger$ that are weak symmetries of the Lindbladian. Therefore, the results of this work can be used to analytically describe the steady states of the model \eqref{eq:dimerized}, including the chirally-symmetric hopping arrangement depicted in Fig. \ref{fig:fig6}a.}

\section{Hidden TRS conditions}
Given a steady-state $\hat{\rho}_\text{ss}$ and corresponding time-reversal operation $\hat{T}$, one can construct the {\it thermofield double state}
\begin{align}
    |\Psi_{\hat{T}}\rangle = \sum_n \sqrt{p_n}|n\rangle_L \hat{T}|n\rangle_R,
\end{align}
where $|n\rangle$ are the eigenvectors of $\hat{\rho}_\text{ss}$, and $p_n$ the corresponding eigenvalues. Sufficient conditions for $\hat{T}$ to be a hidden time-reversal symmetry are \cite{fagnolaGeneratorsKMSSymmetric2010}
\begin{align}
    \Bigg[\bigg(u(\hat{N}_L + \hat{N}_R)-\Delta_\text{eff}\bigg)(\hat{N}_L - \hat{N}_R) + 2\sum_{ij}M_{ij}\hat{\alpha}_{i,+}^\dagger\hat{\alpha}_{j,-}^\dagger\Bigg]|\Psi_{\hat{T}}\rangle  &= 0,\nonumber\\
    \hat{\alpha}_{j,-}|\Psi_{\hat{T}}\rangle&=0,
\end{align}
where $\hat{\alpha}_{j,\pm}\equiv 2^{-1/2}(\hat{a}_{j,L}\pm \hat{a}_{j,R})$. Since the Bogoliubov transformation $\hat{a}_{j,L},\hat{a}_{j,R} \to \hat{\alpha}_{j,+},\hat{\alpha}_{j,-}$ is boson-number preserving, we have $\hat{N}_L+\hat{N}_R = \hat{N}_+ + \hat{N}_-$, where $\hat{N}_\pm :=\sum_j \hat{\alpha}_{j,\pm}^\dagger \hat{\alpha}_{j,\pm}$. Therefore, we can write the above conditions equivalently as
\begin{align}
    \sum_j\hat{\alpha}_{j,-}^\dagger\Bigg[\bigg(u(\hat{N}_+ + \hat{N}_-+1)-\Delta_\text{eff}\bigg)\hat{\alpha}_{j,+} + 2\sum_{i}M_{ij}\hat{\alpha}_{i,+}^\dagger\Bigg]|\Psi_{\hat{T}}\rangle  &= 0,\label{eq:orthosum}\\
    \hat{\alpha}_{j,-}|\Psi_{\hat{T}}\rangle&=0.
\end{align}
Each term in the sum in Eq. \eqref{eq:orthosum} has exactly one boson in the antisymmetric mode $\hat{\alpha}_{j,-}$ and is thus mutually orthogonal with the remaining terms. Therefore, each term in the sum must separately vanish. Factoring the thermofield double into symmetric and antisymmetric components $|\Psi_{\hat{T}}\rangle = |\Psi_+\rangle|\Psi_-\rangle$, we obtain $|\Psi_-\rangle = |0_-\rangle$, i.e. the antisymmetric component is in vacuum, and
\begin{align}
    \bigg((u(\hat{N}_++1) - \Delta_\text{eff})\hat{\alpha}_{j,+}+2\sum_iM_{ij}\hat{\alpha}_{i,+}^\dagger\bigg)|\Psi_+\rangle = 0.\label{eq:sys_of_eqns}
\end{align}

\subsection{Exact solution via representation theory}
\noindent We now proceed to solve the system of equations Eq. \eqref{eq:sys_of_eqns}. To begin, we multiply each constraint on the left by $\hat{\alpha}_{j,+}^\dagger$, yielding
\begin{align}
    \bigg(u(\hat{N}_+ - \Delta_\text{eff})\hat{n}_{j,+}+2\sum_iM_{ij}\hat{\alpha}_{i,+}^\dagger\hat{\alpha}_{j,+}^\dagger\bigg)|\Psi_+\rangle = 0,
\end{align}
where $\hat{n}_{j,+}:=\hat{\alpha}_{j,+}^\dagger \hat{\alpha}_{j,+}$. We then sum over $j$:
\begin{align}
    \bigg(u(\hat{N}_+ - \Delta_\text{eff})\hat{N}_{+}+2\sum_{ij}M_{ij}\hat{\alpha}_{i,+}^\dagger\hat{\alpha}_{j,+}^\dagger\bigg)|\Psi_+\rangle = 0.
\end{align}
Note how the above operator annihilating $|\Psi_{\hat{T}}\rangle$ is almost like the effective nonhermitian Hamiltonian $\hat{H}_\text{eff}:= \hat{H}-i\kappa\hat{N}/2$, except that here the lowering operators associated with the drive are not present. We now identify a "hidden" nonunitary representation of $SU(1,1)$ via
\begin{align}
    \hat{K}_+ &= \frac{1}{2}\sum_{ij}\frac{M_{ij}}{u}\hat{\alpha}_{i,+}^\dagger \hat{\alpha}_{j,+}^\dagger,~~~~\hat{K}_- = \frac{1}{2}\sum_{ij}(uM^{-1})_{ij}\hat{\alpha}_{i,+} \hat{\alpha}_{j,+},~~~~\hat{K}_z=\frac{1}{2}\sum_j\bigg(\hat{\alpha}_{j,+}^\dagger \hat{\alpha}_{j,+}+\frac{1}{2}\bigg).\label{eq:su11_rep}
\end{align}
Note that the representation presented here is unitary if and only if the pairing array is unitary, i.e. $(M/u)^{-1}=(M/u)^\dagger$. In terms of the Lie algebra generators, the hTRS condition simplifies to $\mathcal H_{\text{eff},+}|\Psi_+\rangle = 0$, where
\begin{align}
    \mathcal H_{\text{eff},+} :=\big(\hat{K}_z-N/4-\Delta_\text{eff}/2u\big)\big(\hat{K}_z-N/4\big)+\hat{K}_+.
\end{align}
To solve this condition, we make the ansatz that the dark state inhabits an irreducible representation of $SU(1,1)$, with the vacuum state constituting the vector of lowest weight (with respect to the subalgebra generated by $\hat{K}_z$). This irreducible representation is spanned by vectors of the form
\begin{align}
    |\Psi\rangle =\sum_m c_m\hat{K}_+^m |h_0\rangle,~~~~|h_0\rangle := |\Omega\rangle\label{eq:ansatz}
\end{align}
The use of the notation $h_0$ to denote the bosonic vacuum $|\Omega\rangle:=|0\rangle_L|0\rangle_R$, as well as the fact that the solution Eq. \eqref{eq:ansatz} to the constraints Eq. \eqref{eq:sys_of_eqns} is unique will become apparent in later sections when we review the general representation theory of $SU(1,1)$. For now, however, we will treat Eq. \eqref{eq:ansatz} as an ansatz and proceed to compute the exact coefficients $c_m$ occuring in the expansion:
\begin{align}
    \mathcal H_{\text{eff},+}|\Psi_+\rangle=\sum_{m=1}^\infty\Bigg(m(m -\Delta_\text{eff}/2u)c_m+c_{m-1}\Bigg)\hat{K}_+^m|h_0\rangle = 0,\label{eq:collective_condition}
\end{align}
which yields the solution $c_m\propto(-1)^m/m!(\delta)_m$, where $\delta := 1-\Delta_\text{eff}/2u$, and $(\delta)_m:= \delta(\delta+1)\cdots (\delta+m-1)$ denotes the Pochhammer symbol (rising factorial).

\subsection{Diagonal form of the representation}
By treating the raising operators $\hat{\alpha}_{j,+}^\dagger$ as indeterminates, one can interpret the $SU(1,1)$ raising operator $\hat{K}_+$ presented in Eq. \eqref{eq:su11_rep} as a quadratic form over the complex numbers. Its matrix representation $M/u$ is invariant under transposition and thus admits a kind of singular-value decomposition $M/u = V\Sigma V^T$ (called the Autonne-Takagi factorization \cite{autonneMatricesHypohermitiennesMatrices1915}), with $V$ an $N\times N$ unitary matrix and $\Sigma = \text{diag}(\lambda_1,\dots, \lambda_N)$ the matrix of singular values. Writing $\hat{\beta}_{i,+}^\dagger := \sum_i V_{ij}\hat{\alpha}_{j,+}^\dagger$ for the corresponding basis of singular vectors, we can re-express the $SU(1,1)$ representation \eqref{eq:su11_rep} as follows:
\begin{align}
    \hat{K}_+ &= \frac{1}{2}\sum_{j}\lambda_{j}\hat{\beta}_{j,+}^{\dagger 2},~~~~\hat{K}_- = \frac{1}{2}\sum_{i}\lambda_j^{-1}\hat{\beta}_{j,+}^{2},~~~~\hat{K}_z=\frac{1}{2}\sum_j\bigg(\hat{\beta}_{j,+}^\dagger \hat{\beta}_{j,+}+\frac{1}{2}\bigg).\label{eq:su11_rep_diagonalform}
\end{align}

\section{Exact solution: collective moments}
The simplest quantities to compute in our model are collective observables, that is, moments of normally-ordered monomials in the $SU(1,1)$ algebra.

\subsection{Unitary case}
Our job is easiest when the representation Eq. \eqref{eq:su11_rep} satisfies $\hat{K}_+^\dagger \propto \hat{K}_-$. This happens if and only if the singular values of the matrix $M/u$ are all the same. In this case, the $SU(1,1)$ representation can be made unitary by rescaling $\hat{K}_-$, and the diagonal form of the representation is as follows:
\begin{align}
    \hat{K}_+=\frac{1}{2}\sum_j\lambda \hat{\beta}_{j,+}^{\dagger2},~~~~~~~\hat{K}_-=\frac{1}{2}\sum_j\lambda^{-1} \hat{\beta}_{j,+}^{2},~~~~\hat{K}_z=\frac{1}{2}\sum_j\bigg(\hat{\beta}_{j,+}^\dagger \hat{\beta}_{j,+}+\frac{1}{2}\bigg),
\end{align}
with $\lambda>0$ the unique singular value of the matrix $M/u$. With this, we now turn to compute moments of monomials of $SU(1,1)$ within the representation:
\begin{align}
    \langle\Psi_{\hat{T}}|\hat{K}_+^n \hat{N}_+^k\hat{K}_-^m|\Psi_{\hat{T}}\rangle =  \sum_{p,q}\lambda^{2p} c_p^* c_q\langle h_0|\hat{K}_-^{p}\hat{K}_+^n \hat{N}_+^k \hat{K}_-^m \hat{K}_+^q |h_0\rangle
\end{align}
We reparametrize the sum by defining an auxiliary non-negative integer $l = p-n = q-m$. In terms of $l$:
\begin{align}
    \langle\Psi_{\hat{T}}|\hat{K}_+^n \hat{N}_+^k\hat{K}_-^m|\Psi_{\hat{T}}\rangle &= \sum_{l=0}^\infty \lambda^{2(l+n)} c_{l+n}^*c_{l+m} \langle h_0|\hat{K}_-^l\hat{K}_-^n \hat{K}_+^n\hat{N}_+^k\hat{K}_-^m \hat{K}_+^m \hat{K}_+^l|h_0\rangle\nonumber\\
    & = \lambda^{2n}c_n^*c_m\sum_{l=0}^\infty (2l)^k |c_l|^2\lambda^{2l} \frac{c_{n+l}^*}{c_l^*c_n^*}\frac{c_{m+l}}{c_lc_m}f(n,m,l).
\end{align}
Using the $SU(1,1)$ commutation relations, we evaluate
\begin{align}
    f(n,m,l) &:= \langle h_0|\hat{K}_-^l\hat{K}_-^n \hat{K}_+^n\hat{K}_+^m \hat{K}_-^m \hat{K}_+^l|h_0\rangle\nonumber\\
    &= l! (N/2)_l \frac{(n+l)!}{l!}\frac{(m+l)!}{l!} \frac{(N/2)_{n+l}}{(N/2)_l}\frac{(N/2)_{m+l}}{(N/2)_l} = \frac{(N/2)_n(N/2)_m}{l!}\frac{(n+l)!(m+l)!(N/2+n)_l(N/2+m)_l}{(N/2)_l}.
\end{align}
We then compute
\begin{align}
     (2l)^k |c_l|^2\lambda^{2l} \frac{c_{n+l}^*}{c_l^*c_n^*}\frac{c_{m+l}}{c_lc_m} = |c_l|^2 \lambda^{2l} \frac{l!n!l!m!}{(n+l)!(m+l)!}\frac{(\delta^*)_l(\delta^*)_n(\delta)_l(\delta)_m}{(\delta^*)_{l+n}(\delta^*)_{l+m}} = \frac{\lambda^{2l}n!m!}{(n+l)!(m+l)!(\delta^*+n)_l(\delta+m)_l}.
\end{align}
Putting everything together,
\begin{align}
    \langle\Psi_{\hat{T}}|\hat{K}_+^n\hat{N}_+^k\hat{K}_-^m|\Psi_{\hat{T}}\rangle &=\lambda^{2n} c_n^*c_m (N/2)_n(N/2)_mn!m!\sum_{l=0}^\infty (2l)^k \frac{\lambda^{2l}}{l!}\frac{(N/2+n)_l(N/2+m)_l}{(N/2)_l(\delta^*+n)_l(\delta+m)_l}\nonumber\\
    &= \frac{\lambda^{2n}(-1)^{n+m}(N/2)_n(N/2)_m}{(\delta^*)_n(\delta)_m}\bigg(2z\frac{d}{dz}\bigg)^k \,_2F_3(N/2+n,N/2+m; N/2,\delta^*+n,\delta+m; z)\Bigg|_{z = \lambda^2},\label{eq:collective}
\end{align}
where here, $\,_pF_q$ denotes the generalized hypergeometric function, defined as the analytic extension of the following infinite series:
\begin{align}
    \,_pF_q(a_1,\dots a_p;b_1,\dots b_q;z)= \sum_{l=0}^\infty \frac{\prod_{m=1}^p(a_m)_l}{\prod_{m=1}^q(b_m)_l}\frac{z^l}{l!}.
\end{align}
Up to this point, we have computed $|\Psi_{\hat{T}}\rangle$ up to a normalization prefactor. This normalization can also be computed in closed form: $\langle \Psi_{\hat{T}}|\Psi_{\hat{T}}\rangle = \,_1F_2(N/2;\delta,\delta^*;\lambda^2)$, thus completing the characterization of collective moments in the case that the representation \eqref{eq:su11_rep} is unitary.

\subsection{Nonunitary case}
\noindent We now accomplish the same task, but in the more general regime where the singular values $\lambda_j$ are possibly non-degenerate. In this situation, there is less structure to the calculation, but moments of $SU(1,1)$ monomials may nonetheless be extracted efficiently:
\begin{align}
    \langle \Psi_{\hat{T}}|\hat{K}_+^n\hat{N}_+^k \hat{K}_-^m|\Psi_{\hat{T}}\rangle&= \sum_{p,q}^\infty c_p^* c_q \langle h_0|\hat{K}_+^{\dagger p}\hat{K}_+^n \hat{N}_+^k \hat{K}_-^m \hat{K}_+^q|h_0\rangle
\end{align}
We reparametrize the sum by defining an auxiliary non-negative integer $l = p-n = q-m$. In terms of $l$:
\begin{align}
     \langle \Psi_{\hat{T}}|\hat{K}_+^n\hat{N}_+^k \hat{K}_-^m|\Psi_{\hat{T}}\rangle&= \sum_{l=0}^\infty c_{l+n}^* c_{l+m} (2l)^k \langle h_0|\hat{K}_+^{\dagger l}\hat{K}_+^{\dagger n} \hat{K}_+^n \hat{K}_-^m\hat{K}_+^m \hat{K}_+^l|h_0\rangle\nonumber\\
    & = c_n^*c_m\sum_{l=0}^\infty |c_l|^2 \frac{c_{l+n}^*}{c_l^*c_n^*} \frac{c_{l+m}}{c_lc_m} (2l)^k g(n,m,l).
\end{align}
We evaluate:
\begin{align}
    g(n,m,l) &= \frac{(N/2)_{m+l}(m+l)!}{(N/2)_ll!}\langle h_0|\hat{K}_+^{\dagger (n+l)}\hat{K}_+^{n+l}|h_0\rangle:= \frac{(N/2)_{m+l}(m+l)!(n+l)!(n+l)!}{(N/2)_ll!}\Phi_{n+l}(\vec{\lambda};N),
\end{align}
where $\Phi_n(\vec{\lambda};N):=(n!)^2\langle h_0| \hat{K}_+^{\dagger n}\hat{K}_+^n|h_0\rangle$ is a form factor to be evaluated later on. Therefore, in total we have
\begin{align}
    \langle \Psi_{\hat{T}}|\hat{K}_+^n\hat{N}_+^k\hat{K}_-^m|\Psi_{\hat{T}}\rangle = \frac{(-1)^{n+m}(N/2)_m}{(\delta^*)_n(\delta)_m} \sum_{l=0}^\infty \frac{(N/2+m)_l(n+l)!(2l)^k}{(\delta^*+n)_l(\delta+m)_l(N/2)_ll!}\Phi_{n+l}(\vec{\lambda};N).
\end{align}
All that is left is to compute the normalization, which comes out to $\langle \Psi_{\hat{T}}|\Psi_{\hat{T}}\rangle = \sum_{l=0}^\infty \Phi_l(\vec{\lambda};N)/(\delta)_l(\delta^*)_l$. Up until this point, we have given formulae in terms of the form factors $\Phi_n(\vec{\lambda};N):=\langle h_0| \hat{K}_+^{\dagger n}\hat{K}_+^n|h_0\rangle/(n!)^2$, which, in the nonunitary case, cannot be evaluated purely using the internal relations of the $SU(1,1)$ algebra. Instead, via the multinomial expansion, we are left with a sum over an exponential number of terms:
\begin{align}
    \Phi_n(\vec{\lambda};N) = \frac{1}{n!^2}\langle h_0|\Bigg(\frac{1}{2}\sum_j \lambda_j\hat{\beta}_{j,+}^{ 2}\Bigg)^n\Bigg(\frac{1}{2}\sum_j \lambda_j\hat{\beta}_{j,+}^{\dagger 2}\Bigg)^n|h_0\rangle = \sum_{\sum_j k_j = n}\prod_j (1/2)_{k_j}\lambda_j^{2k_j}
\end{align}
Our work would not be complete if we did not give an efficient prescription for evaluating these sums. The key to evaluating these sums efficiently is the following observation:
\begin{align}
    \Phi_n(\vec{\lambda};N) = \sum_{p=0}^n(1/2)_p \lambda_N^{2p}\sum_{\sum_{j=1}^{N-1} k_j = n-p}\prod_j (1/2)_{k_j}\lambda_j^{2k_j} = \sum_{p=0}^n(1/2)_p \lambda_N^{2p} \Phi_{n-p}(\vec{\lambda};N-1)\label{eq:combinatorics}
\end{align}
We note that the above constitutes a recursion relation for $\Phi_n(\vec{\lambda};N)$ with boundary condition $\Phi_p(\vec{\lambda};1) = (1/2)_p \lambda_1^{2p}$. Note that via this recursion relation, and fixing some boson-number cutoff $k$, the collection $\{\Phi_p(\vec{\lambda};N)\}_{p\leq k}$ of form factors may be evaluated using only $O(k^2N)$ floating-point operations.

\section{Exact solution: local moments}
We now turn to a more difficult task: that of computing (equal-time) correlation functions of local observables, that is, observables that are not collective in nature. To accomplish this efficiently in the unitary case, we must solve the "addition of angular momentum" problem for $SU(1,1)$, i.e. we must understand how to decompose a tensor product of $SU(1,1)$ representations into irreducible components. Luckily, this task is easily solved in terms of the theory of harmonic functions on $\mathbb{R}^N$. In the nonunitary case, however, the $SU(1,1)$ structure is not so useful, and instead some generalization of the combinatorics in \eqref{eq:combinatorics} is needed to compute observables.

To make our task easier, we will compute observables in the basis of modes $\hat{b}_i^\dagger := \sum_{j}V_{ij}\hat{a}_j^\dagger$ that diagonalizes the pair-driving matrix $M/u$. 

\subsection{Addition of angular momentum for $SU(1,1)$}
We can write our global $SU(1,1)$ representation \eqref{eq:su11_rep_diagonalform} as a tensor product of local representations:
\begin{align}
    \hat{K}_+^{(j)} \equiv \frac{1}{2}\lambda_j\hat{\beta}_{j,+}^{\dagger2},~~~~\hat{K}_-^{(j)} \equiv \frac{1}{2}\lambda^{-1}_j\hat{\beta}_{j,+}^{2},~~~~\hat{K}_z^{(j)} \equiv \frac{1}{2}\bigg(\hat{\beta}_{j,+}^{\dagger}\hat{\beta}_{j,+} + \frac{1}{2}\bigg)\label{eq:localreps}
\end{align}
It is easy to see that each local representation is reducible, and has the decomposition $V^{(j)}  = V^{(j)}_+ \oplus V^{(j)}_-$, where $V^{(j)}_\pm$ denotes the subspace consisting of all states with a fixed boson-number parity $\pm 1$.  Our global $SU(1,1)$ representation $\hat{K}_+ , \hat{K}_-,\hat{K}_z$ is the $N$-fold tensor product $\otimes_j V^{(j)}$, and has the following decomposition into irreducible subrepresentations:
\begin{align}
    \otimes_j V^{(j)}  \simeq \bigoplus_{l=0}^\infty\Bigg(\bigoplus_{p=1}^{d_l} V_l^{(p)}\Bigg), \label{eq:the_decomp}
\end{align}
where each irreducible subrepresentation $V_l^{(p)}$ takes the form
\begin{align}
    |\Psi_l^{(p)}\rangle = \sum_{m=0}^\infty c_m \hat{K}_+^m |h_{l}^{(p)}\rangle,
\end{align}
and $|h_l^{(1)}\rangle,\dots,|h_l^{(d_l)}\rangle$ is some orthonormal basis of the subspace of the kernel of $\hat{K}_-$ consisting of those states with fixed total photon number equal to $l$. Later on we sketch a proof of the decomposition \eqref{eq:the_decomp} using the Segal-Bargmann representation \cite{bargmann_hilbert_1961,bargmann_hilbert_1967}, which represents bosonic states as multivariate analytic functions, and bosonic creation- and annihilation operators as partial differential operators. In the Segal-Bargmann representation, the $SU(1,1)$ vacua $|h_l^{(p)}\rangle$ are represented as homogeneous harmonic polynomials of degree $l$, hence the notation "$h_l$".

\subsection{Unitary case}
In the unitary case, the decomposition \eqref{eq:the_decomp} becomes an orthogonal direct-sum decomposition, with the superselection rules $\langle h_l^{(m)}|\hat{K}_+^{\dagger p}\hat{K}_+^q|h_{l'}^{(m')}\rangle = \delta_{l,l'}\delta_{m,m'} \delta_{p,q} p! \lambda^{2p}(N/2+ l)_p$. These simple rules allow us to compute any local quantity of interest in our model. For the case of quadratic correlation functions, however, such rules are not necessary. To see this, it suffices to use the weak permutation symmetry $\hat{b}_i\leftrightarrow \hat{b}_j$, as well as the weak parity symmetry $\hat{b}_i \to -\hat{b}_i$ of the Lindbladian $\mathcal L$:
\begin{align}
    \langle \hat{b}_{i}^\dagger\hat{b}_{i+r}\rangle &= \delta_{r,0} \frac{1}{N}\sum_{j}\langle \hat{b}_j^\dagger\hat{b}_j\rangle  = \frac{\delta_{r,0}}{N}\langle \Psi_{\hat{T}}|\hat{N}_+|\Psi_{\hat{T}}\rangle,~~~~~~\langle \hat{b}_{i}\hat{b}_{i+r}\rangle = \delta_{r,0} \frac{1}{N}\sum_{j}\langle \hat{b}_j^2\rangle  = \frac{\delta_{r,0}}{N}\langle \Psi_{\hat{T}}|\hat{K}_-|\Psi_{\hat{T}}\rangle
\end{align}
Higher-order correlations, are not expressible in terms of collective moments, and so in general the decomposition \eqref{eq:the_decomp}, along with the associated superselection rules, serve as a useful guide. As an example, we compute the pair-correlation function, as well as the density-density correlation function, and leave the general case to the reader. We compute the pair correlation function first:
\begin{align}
    \langle \hat{b}_i^{\dagger2}\hat{b}_{i+r}^2\rangle &= \frac{1}{4}\langle \Psi_{\hat{T}}|\hat{\beta}_{i+r,+}^{2}\hat{\beta}_{i,+}^{\dagger 2}|\Psi_{\hat{T}}\rangle-\frac{2\delta_{r,0}}{N}\langle \Psi_{\hat{T}}|\hat{K}_z|\Psi_{\hat{T}}\rangle=\frac{1}{4}\sum_{p,q}c_p^*c_q \langle h_0|\hat{\beta}_{i,+}^{2}\hat{K}_+^{\dagger p}\hat{K}_+^q\hat{\beta}_{i+r,+}^{\dagger 2}|h_0 \rangle-\frac{2\delta_{r,0}}{N}\langle \Psi_{\hat{T}}|\hat{K}_z|\Psi_{\hat{T}}\rangle
\end{align}
For the calculation to proceed, it is necessary to decompose the states $\hat{\beta}_{i,+}^{\dagger 2}|h_0 \rangle$ into $SU(1,1)$ vacua. Although this can be done by hand in this simple case, in general it is useful to automate this process (especially for higher-order observables). For this purpose, the HFT {\it Mathematica} package \cite{axlersheldonHFT2020} is especially relevant, in particular the function $\text{harmonicDecomposition}[]$, which automatically extracts the decomposition
\begin{align}
    \hat{\beta}_{i,+}^{\dagger2}|h_0\rangle = \sqrt{\frac{2N-2}{N}}|h_2^{(i)}\rangle  + \frac{2\hat{K}_+}{\lambda N}|h_0\rangle,
\end{align}
where $|h_2^{(i)}\rangle := \frac{1}{\lambda}\sqrt{\frac{2N}{N-1}}(\hat{K}_+^{(i)} - \frac{\hat{K}_+}{N})|h_0\rangle$ generates an irreducible subrepresentation with weight $N/4 +1$. We thus obtain:
\begin{align}
    \frac{1}{4}\sum_{p,q}c_p^*c_q \langle h_0|\hat{\beta}_{i,+}^{2}\hat{K}_+^{\dagger p}\hat{K}_+^q\hat{\beta}_{i+r,+}^{\dagger 2}|h_0 \rangle&= \frac{2N-2}{N^3\lambda^2}\sum_{l} |c_l|^2 \langle h_0|\hat{K}_+^{\dagger l+1} \hat{K}_+^{l+1}|h_0\rangle+\frac{N-1}{2N}\sum_l|c_l|^2\langle h_2^{(i)}|\hat{K}_+^{\dagger l}\hat{K}_+^l|h_2^{(i+r)}\rangle \nonumber\\
    &= \Bigg(\frac{2N-2}{N^3} \bigg[z\frac{\partial}{\partial z} + 1\bigg]\,_1F_2(N/2;\delta,\delta^*;z) + \frac{N\delta_{r,0}-1}{2N} \,_1F_2(N/2+2;\delta,\delta^*;z)\Bigg)\Bigg|_{z=\lambda^2}.
\end{align}
We now compute the density-density correlation function $\langle \hat{b}_i^\dagger \hat{b}_i\hat{b}_{i+r}^\dagger \hat{b}_{i+r}\rangle$. Since the case $r=0$ is subsumed by the previous calculation, without loss of generality we can assume $r\neq 0$. Again, we split the calculation into a collective part and a noncollective part (that populates a higher weight representation):
\begin{align}
    \langle \hat{b}_i^\dagger \hat{b}_i\hat{b}_{i+r}^\dagger \hat{b}_{i+r}\rangle = \frac{1}{4}\sum_{p,q}c_p^*c_q\langle h_0|\hat{\beta}_i\hat{\beta}_{i+r}\hat{K}_+^{\dagger p} \hat{K}_+^q\hat{\beta}_i^\dagger\hat{\beta}_{i+r}^\dagger|h_0\rangle - \frac{1}{N}\langle \Psi_{\hat{T}}| \hat{K}_z|\Psi_{\hat{T}}\rangle.
\end{align}
We evaluate the noncollective part, by noticing that $|h_2^{(i,j)}\rangle:=\hat{\beta}_{i,+}^\dagger\hat{\beta}_{j,+}^\dagger |h_0\rangle$ generates an irreducible subrepresentation with weight $N/4 + 1$:
\begin{align}
    \frac{1}{4}\sum_{p,q}c_p^*c_q\langle h_0|\hat{\beta}_i\hat{\beta}_{i+r}\hat{K}_+^{\dagger p} \hat{K}_+^q\hat{\beta}_i^\dagger\hat{\beta}_{i+r}^\dagger|h_0\rangle &= \frac{1}{4}\sum_l |c_l|^2\langle h_2^{(i,i+r)}|\hat{K}_+^{\dagger l} \hat{K}_+^l|h_2^{(i,i+r)}\rangle=\frac{1}{4}\,_1F_2(N/2+2;\delta,\delta^*;\lambda^2).
\end{align}

\subsubsection*{Higher-order moments}
The general procedure for calculating higher order moments is no different from what we have done in the previous subsection: given a higher-order moment to be evaluated, one writes the expectation value in terms of an antinormally-ordered correlation function involving the mode operators $\hat{\beta}_{j,+},\hat{\beta}_{j,+}^\dagger$, and then uses standard harmonic analysis software \cite{axlersheldonHFT2020} to decompose the resulting states into components lying in higher-weight subrepresentations. Iterating this process yields formulae in terms of generalized hypergeometric functions and their derivatives.

\subsection{Nonunitary case}
We now turn to the most general task of computing local correlation functions in our global Bose-Hubbard model, in the nonunitary regime. In this case the calculation has the least amount of structure, and so here we just present the most general result. To better organize the calculation, we will write the purification $|\Psi_{\hat{T}}\rangle$ as a power series
\begin{align}
    |\Psi_{\hat{T}}\rangle = \sum_{\vec{m}\in \mathbb{N}^N} \frac{c_{\vec{m}}}{\vec{m}!} \prod_{j}\hat{\beta}_{j,+}^{\dagger m_j}|h_0\rangle,
\end{align}
with $c_{\vec{m}}=\prod_j(4\lambda_j)^{m_j}(1/2)_{m_j}/(\delta)_{\sum_j m_j}$. From the above form of the purification, we can calculate the parametric form of any normally-ordered correlation function:
\begin{align}
    \langle \hat{b}_1^{\dagger n_1}&\cdots  \hat{b}_N^{\dagger n_N}\hat{b}_1^{m_1}\cdots  \hat{b}_N^{m_N}\rangle=\frac{1}{\sqrt{2^{\sum_jn_j+m_j}}}\langle \Psi_{\hat{T}}|\hat{\beta}_{1,+}^{\dagger n_1}\cdots  \hat{\beta}_{N,+}^{\dagger n_N}\hat{\beta}_{1,+}^{m_1}\cdots  \hat{\beta}_{N,+}^{m_N}|\Psi_{\hat{T}}\rangle=\frac{1}{\sqrt{2^{\sum_jn_j+m_j}}} \sum_{\vec{k}\in \mathbb{N}^N}\frac{c^*_{\vec{n}+\vec{k}}c_{\vec{m}+\vec{k}}}{\vec{k}!} \label{eq:useless_series}
\end{align}
There are a number of issues, however, with the series expression given above: firstly, due to the weak symmetry $\hat{b}_j\to -\hat{b}_j$ of $\mathcal L$, only correlation functions with $n_j\equiv m_j$ modulo two are nonzero. Secondly, the series expression naively seems to be useless, as, for a fixed total boson number cutoff, the series contains a number terms that is exponentially growing with $N$. Therefore, the naive way of evaluating \eqref{eq:useless_series} scales no better than a direct simulation of the original master equation.

We will resolve both of these issues now: first of all, we can efficiently parametrize all of the nonzero correlators by replacing $\vec{n} \to 2\vec{n}+\vec{b},\vec{m}\to 2\vec{m} + \vec{b}$, where $\vec{b}\in \mathbb{F}_2^N$ is a fixed vector of booleans. To exponentially reduce the complexity of summing the series \eqref{eq:useless_series}, we define generalized combinatorial form factors
\begin{align}
    \Phi_l(\vec{\lambda},\vec{n},\vec{m},\vec{b};N) := \sum_{\sum_j k_j = l}\prod_j\frac{(1/2 + n_j + b_j)_{k_j}(1/2 + m_j + b_j)_{k_j}}{(2k_j+b_j)!}(4\lambda_j)^{2k_j}.\label{eq:theformfactors}
\end{align}
Indeed, in terms of the above form factors, the series expressions for normally-ordered moments simplify quite considerably:
\begin{align}
    \langle& \hat{b}^{\dagger 2\vec{n}+\vec{b}}\hat{b}^{ 2\vec{m}+\vec{b}}\rangle  =\frac{1}{\sqrt{2^{\sum_jn_j+m_j+b_j}}} \sum_{\vec{k}\in \mathbb{N}^N}\frac{c^*_{2\vec{n}+\vec{b} +\vec{k}}c_{2\vec{m}+\vec{b} +\vec{k}}}{\vec{k}!}\nonumber\\
    &=\frac{1}{\sqrt{2^{\sum_jn_j+m_j+b_j}}} \sum_{\vec{k}\in \mathbb{N}^N}\frac{c^*_{2(\vec{n}+\vec{b} +\vec{k})}c_{2(\vec{m}+\vec{b} +\vec{k})}}{(2\vec{k}+\vec{b})!}=  \frac{c_{2(\vec{n}+\vec{b})}^*c_{2(\vec{m}+\vec{b})}}{\sqrt{2^{\sum_jn_j+m_j+b_j}}}\sum_{l=0}^\infty \frac{\Phi_l(\vec{\lambda},\vec{n},\vec{m},\vec{b})}{(\delta^*+ \sum_j n_j + \sum_jb_j)_l(\delta+ \sum_j m_j + \sum_jb_j)_l}. \label{eq:thefullexpression}
\end{align}
Our job would not be finished if we did not give an efficient prescription for evaluating the form factors $\Phi_l$. Our task is made easier, however, by observing that, when $\vec{n}=\vec{m}=\vec{b} = \vec{0}$, the form factors \eqref{eq:theformfactors} reduce to the form factor $\Phi_n(\vec{\lambda};N)$ used previously to compute collective moments. In fact, these more general form factors satisfy an analogous recursion relation:
\begin{align}
    \Phi_l(\vec{\lambda},\vec{n},\vec{m},\vec{b};N) &= \sum_{p=0}^n\frac{(1/2+n_N+b_N)_p(1/2+m_N+b_N)_p}{(2p+b_N)!} (4\lambda_N)^{2p} \Phi_{n-p}(\vec{\lambda},\vec{n},\vec{m},;N-1).\label{eq:combinatorics2}
\end{align}
We note that the above recursion relation has the boundary condition $\Phi_p(\vec{\lambda},\vec{n},\vec{m},\vec{b};1) = (1/2+n_1+b_1)_p(1/2+m_1+b_1)_p(4\lambda_1)^{2p}/(2p+b_1)!$. Therefore, the following $k$th-order approximant for each normally-ordered moment,
\begin{align}
    \frac{c_{2(\vec{n}+\vec{b})}^*c_{2(\vec{m}+\vec{b})}}{\sqrt{2^{\sum_jn_j+m_j+b_j}}}\sum_{l=0}^k \frac{\Phi_l(\vec{\lambda},\vec{n},\vec{m},\vec{b})}{(\delta^*+ \sum_j n_j + \sum_jb_j)_l(\delta+ \sum_j m_j + \sum_jb_j)_l},\label{eq:thefullsum}
\end{align}
may be evaluated using only $O(k^2N)$ operations. We now factor in considerations as to the scaling of the total-particle-number cutoff $k$. Assuming the onsite density $\bar{n}$ converges as $N\to\infty$, the total number of particles is $O(N)$, and so the cutoff typically scales with the system size: $k \sim O(N)$. Therefore, factoring in all considerations, time-complexity of evaluating a single normally-ordered moment is roughly $O(N^3)$.

\subsubsection*{Quadratic correlation functions}
Up until this point, we have given expressions for normally-ordered moments in the basis of singular modes $\hat{b}_i^\dagger := V_{ij}\hat{a}_j^\dagger$. To obtain correlation functions in the spatial mode basis, we must expand each $\hat{a}$-mode in terms of $\hat{b}$-modes:
\begin{align}
    \langle \hat{a}_i^\dagger\hat{a}_{i+r}\rangle &= \sum_{j}V_{i,j}V_{i+r,j}^*\langle\hat{b}_j^\dagger \hat{b}_j\rangle,~~~~~~~~~\langle \hat{a}_i\hat{a}_{i+r}\rangle = \sum_{j}V_{i,j}^*V_{i+r,j}^*\langle\hat{b}_j^2\rangle 
\end{align}
In both cases, one has to evaluate $N$ normally-ordered moments in the $\hat{b}$-basis. By the previous estimates, each normally-ordered moment takes $O(N^3)$ floating-point operations to evaluate, so that in general it takes $O(N^4)$ floating-point operations to evaluate a quadratic correlation function.

\subsubsection*{Quartic correlation functions}
We now turn to evaluate quartic correlators using the same re-expansion technique. We first compute the pair-correlation function:
\begin{align}
    \langle \hat{a}_i^{\dagger 2}\hat{a}_{i+r}^2\rangle = \sum_{j_1,j_2,j_3,j_4}V_{i,j_1}V_{i,j_2}V_{i+r,j_3}^*V_{i+r,j_4}^*\langle \hat{b}_{j_1}^\dagger\hat{b}_{j_2}^\dagger\hat{b}_{j_3}\hat{b}_{j_4}\rangle =  2\sum_{j\neq j'}V_{i,j}V_{i,j'}V_{i+r,j}^*V_{i+r,j'}^*\langle \hat{b}_j^\dagger\hat{b}_{j'}^\dagger\hat{b}_j\hat{b}_{j'}\rangle+\sum_{j,j'}V_{i,j}^2V_{i+r,j'}^{*2}\langle \hat{b}_j^{\dagger 2}\hat{b}_{j'}^2\rangle.
\end{align}
The calculation for the density-density correlation function (assuming $r\neq 0$) is very similar:
\begin{align}
    \langle \hat{n}_i\hat{n}_{i+r}\rangle&= \sum_{j_1,j_2,j_3,j_4}V_{i,j_1}V_{i+r,j_2}V_{i,j_3}^*V_{i+r,j_4}^*\langle \hat{b}_{j_1}^\dagger\hat{b}_{j_2}^\dagger\hat{b}_{j_3}\hat{b}_{j_4}\rangle\nonumber\\
    &=\frac{1}{2}\sum_{j\neq j'}|V_{i,j}V_{i+r,j'}+V_{i+r,j}V_{i,j'}|^2\langle \hat{b}_j^\dagger\hat{b}_{j'}^\dagger\hat{b}_j\hat{b}_{j'}\rangle+\sum_{j,j'}V_{i,j}V_{i+r,j}V_{i,j'}^*V_{i+r,j'}^*\langle \hat{b}_j^{\dagger 2}\hat{b}_{j'}^2\rangle.
\end{align}
In both cases, one has to evaluate $O(N^2)$ normally-ordered moments in the $\hat{b}$-basis. By the previous estimates, each normally-ordered moment takes $O(N^3)$ floating-point operations to evaluate, so that in general it takes $O(N^5)$ floating-point operations to evaluate a quartic correlation function.

$~$

\section{Phenomenology of the exact solution}

\subsection{$SU(1,1)$ coherent states}

\begin{figure}
    \centering
    \includegraphics[width = 0.9\columnwidth]{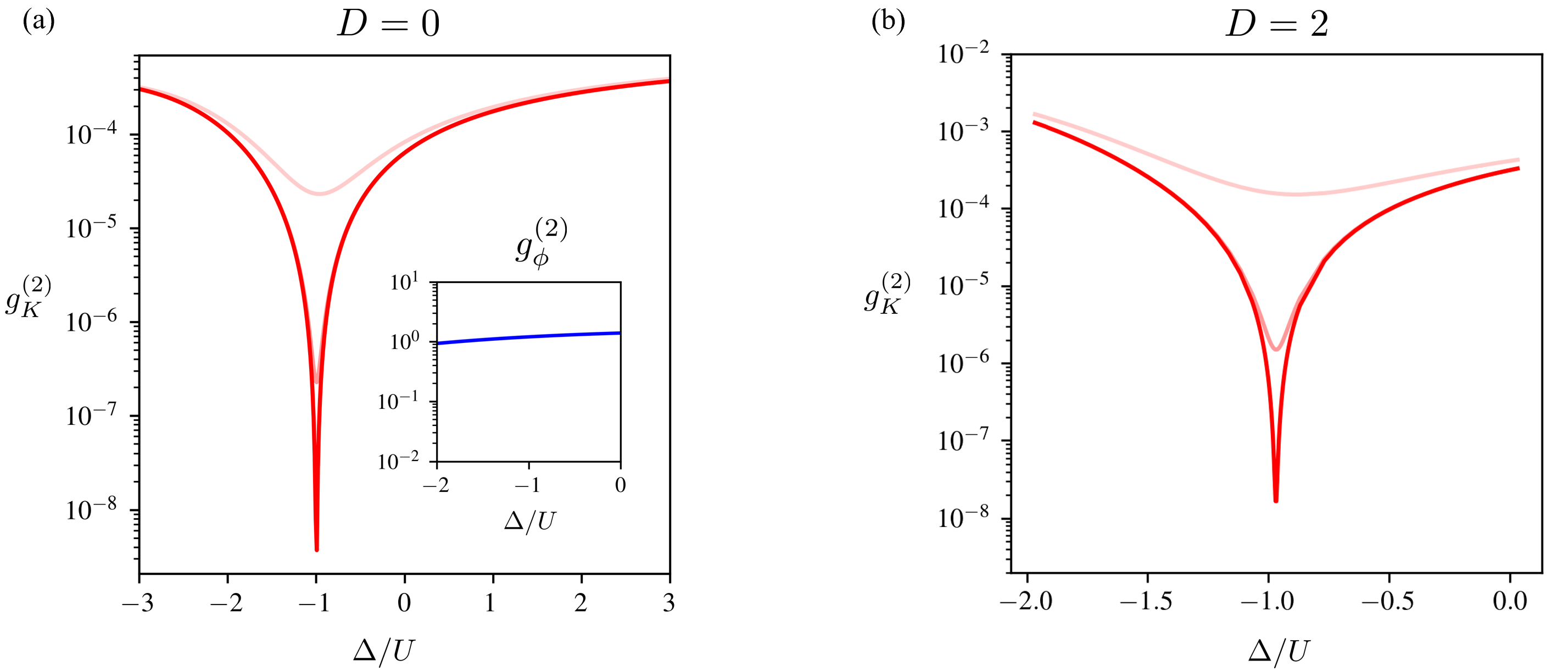}
    \caption{{\bf Bosonic pairing fluctuations near the PCS regime}. (a) Here, we plot the normalized fluctuations $g^{(2)}_K := (\langle \hat{k}_-^\dagger \hat{k}_-\rangle - |K|^2)/|K|^2$ in the nonlocal pairing observable $K:= \langle \hat{k}_-\rangle$, for different values of loss $\kappa \in \{0.01U, 0.1U, U\}$ (red curves; transparency increases with increasing loss). Note the sharp dip exactly at $\Delta_\text{PCS} = U(2-N)/N$. Here, $G = U, \Lambda = 0,N=500$. Inset: We plot the normalized fluctuations $g^{(2)}_\phi := (\langle \hat{a}^{\dagger 2} \hat{a}^2\rangle - |\phi|^2)/|\phi|^2$ in the local onsite pairing $\phi := \langle \hat{a}_j^2\rangle$. Parameters same as before. (b) Same as panel (a), but for $D=2$ with  $N=8\times 8$ and periodic boundary conditions. Here, $\Lambda = 2U = 4G$.}
    \label{fig:my_label}
\end{figure}

When $\delta=N/2$, the purification $|\Psi_{\hat{T}}\rangle$ is an $SU(1,1)$-coherent state in the sense of \cite{barutNewCoherentStates1971}, that is, an eigenstate of the lowering operator $\hat{K}_-$. This holds regardless of whether the representation is unitary or not:
\begin{align}
    \hat{K}_-|\Psi_{\hat{T}}\rangle = \sum_{m=0}^\infty \frac{(-1)^m}{(N/2)_m} \frac{\hat{K}_-\hat{K}_+^m}{m!} |h_0\rangle = \sum_{m=0}^\infty \frac{(-1)^{m+1}}{(N/2)_{m+1}}(m+1)(N/2+m) \frac{\hat{K}_+^m}{m!} |h_0\rangle = -|\Psi_{\hat{T}}\rangle.
\end{align}
The above identity has a dramatic effect on the physical steady state $\hat{\rho}_\text{ss}$. In particular, fluctuations in $K:=\langle \hat{k}_-\rangle$ exactly vanish when $\delta=N/2$, whereas fluctuations in the onsite pairing $\phi:= \langle \hat{a}_j^2\rangle$ remain nonzero and show no special behavior. Here, $\hat{k}_-:= \frac{1}{2}\sum_{ij}(uM^{-1})_{ij}\hat{a}_i \hat{a}_j$ is a pair-lowering operator involving only the physical lattice modes $\hat{a}_j$.

\subsection{DMFT analysis of the $D=0$ model}

\subsubsection*{DMFT expansion}
We now use dynamical mean field theory (DMFT) to derive a large-$N$ expansion for our $D=0$ model. To derive this expansion, we first fix a site (labelled $j=0$) in our lattice model, and attempt to integrate-out the remaining degrees of freedom. Typically this integration procedure is carried out within the Schwinger-Keldysh formalism, which states that the action for the full $D = 0$ lattice model is
\begin{align}
    S &= \int dt  \sum_{\sigma = \pm 1} \sigma \bigg[\sum_j(\alpha_{j,\sigma} \partial_t\alpha_{j,\sigma}^*-\Delta n_{j,\sigma} + G(\alpha_{j,\sigma}^2+ c.c.))+\frac{U}{N}\bigg(\sum_j n_{j,\sigma}\bigg)^2\bigg]\nonumber\\
    &~~~~~~~~~~~~~~~~~~~~~~~~~~~~~~~~~~~~~~~~~~~~~~~~~~+i\kappa \int dt \sum_j \bigg(\alpha_{j,+}\alpha_{j,-} - \frac{1}{2}\sum_{\sigma = \pm 1}|\alpha_{j,\sigma}|^2\bigg),~~~~~~~~~~~~~~~~n_{j,\sigma} :=|\alpha_{j,\sigma}|^2
\end{align}
where $\alpha_{j,\pm}=\alpha_{j,\pm}(t)$ are complex fields associated to the bosonic degrees of freedom $\hat{a}_j,\hat{a}_j^\dagger$, as is standard in the Keldysh formalism. Within this formalism, the goal is now to compute the effective action for a fixed site $j=0$. One can compute a large-$N$ asymptotic expansion for the effective action for the fields $\alpha_{0,+},\alpha_{0,-}$, which takes the following self-consistent form at leading-order:
\begin{align}
    S_\text{eff} = S_\text{free} + \int dt \sum_{\sigma = \pm 1}2\sigma U n_{0,\sigma} \langle n_{0,\sigma}\rangle_\text{eff}+ O(N^{-1}),\label{eq:Seff}
\end{align}
where $S_\text{free}$ is the Keldysh action for a single site, without the Bose-Hubbard interaction, and $\langle \cdot\rangle_\text{eff}$ denotes an average taken with respect to the effective action. Note that this leading-order theory is Markovian. Therefore, as $N\to\infty$, one can evolve observables for a fixed site using the self-consistent Lindbladian
\begin{align}
    \mathcal L_\text{eff}\hat{\rho}_0 = -i[\Delta_\text{eff}(\hat{\rho}_0) \hat{n}_0 + G\hat{a}_0^{\dagger2}+h.c., \hat{\rho}_0] + \kappa\mathcal D[\hat{a}_0]\hat{\rho}_0,~~~~~~~\Delta_\text{eff}(\hat{\rho}_0) =\Delta+2U\text{Tr}[\hat{\rho}_0\hat{n}_0].
\end{align}
One can then solve for the steady-state density $\bar{n}\equiv\bar{n}^\text{MF}$ within this leading-order mean-field description, leading to a cubic equation for $\bar{n}^\text{MF}$. This leads to a large-$N$ phase diagram where regions exist with up to three self-consistent solutions for the density. We obtain good agreement between this leading-order DMFT description and the exact solution, and find that

\begin{figure}
    \centering
    \includegraphics[width = \columnwidth]{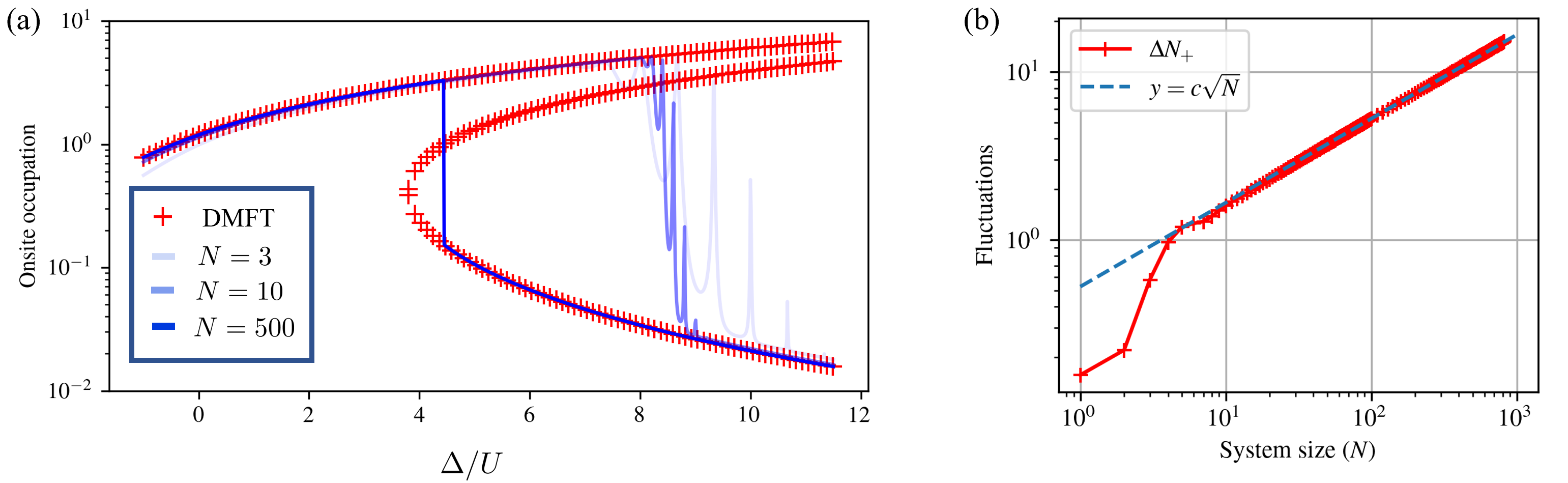}
    \caption{{\bf Benchmarking DMFT using the exact solution (in $D = 0$)}. (a) Here, we plot the mean onsite occupation $\bar{n}$ as a function of detuning, for $G = U,\kappa = 0.01U$. Note that we obtain asymptotic agreement with DMFT in the limit that $N\to\infty$. {To see work where a similar kind of mean field theory was benchmarked by exact diagonalization results in a permutation-symmetric spin model, see \cite{ising_transitions2021}}. (b) Here, we plot the rms fluctuations in $\hat{N}_+$ using the exact solution. We observe the empirical scaling $\Delta N_+ :=\sqrt{\langle \hat{N}_+^2\rangle-\langle \hat{N}_+\rangle^2} =O(N^{1/2})$ so that $\Delta N_+/N$ vanishes as $N\to\infty$, ensuring the asymptotic convergence of the wavefunction $|\Psi_{\hat{T}}\rangle$ of the paired boson gas to the form predicted by DMFT. Here, $G=U/10,\kappa = U/100$, and $\Delta = 0$.}
    \label{fig:my_label}
\end{figure}

\begin{itemize}
    \item  The location $\Delta_c$ of the 1st-order phase transition obtained from the exact solution
    approaches the location of the bifurcation in the MFT cubic self-consistency condition as $N\to\infty$. However, this convergence is extremely slow (finite-size effects are still noticeable for $N\sim 10^3$).
    \item The tristable region of the mean-field phase diagram has a unique point with maximal loss rate $\kappa_*$. Let $(\Delta_*,\kappa_*)$ denote this point. We call this point the {\it mean-field critical-point}. We find that {\it the mean-field critical point is precisely the same as the critical point} in the phase diagram obtained from the exact solution, that is,
    $$\kappa_*=\kappa_c,~~~~~~~\Delta_c(\kappa_c) = \Delta_*.$$
\end{itemize}

\subsubsection*{Establishing the validity of DMFT}
Normally DMFT expansions are hard to rigorously justify directly, even in the limit of large coordination number $z\to\infty$. We will nonetheless give the standard justification here, and then see how the exact hTRS solution yields a much more direct perspective, at least with respect to the steady state problem. The usual justification for the asymptotic result \eqref{eq:Seff} is as follows: note that we can view our Hubbard interaction as a general (extended) Bose-Hubbard interaction for a general graph $\mathcal G$ with $\mathcal G= K_N$, i.e. a complete graph,
\begin{align}
    \frac{U}{N}\bigg(\sum_j \hat{n}_j\bigg)^2 = \frac{U}{z}\sum_{\langle i,j\rangle\in \mathcal G}\hat{n}_i\hat{n}_j\Bigg|_{\mathcal G = K_N}.
\end{align} 
The large-$N$ asymptotic result \eqref{eq:Seff} can be established, by copying exactly the calculation in \cite{scarlatellaDynamicalMeanFieldTheory2021, strandNonequilibriumDynamicalMeanField2015}, but by performing the cumulant expansion therein with respect to the density fields $n_{j,\sigma}$ instead of the ordinary complex fields $\alpha_{j,\sigma}$ (the calculation is relatively unilluminating, and for the sake of brevity, we omit it here). This has the effect of establishing the desired result for $\mathcal G$ a regular tree graph with coordination number $N$, instead of a complete graph. After performing such a calculation, one then waves one's hands and claims that the same asymptotics \eqref{eq:Seff} holds on a complete graph.

The exact solution yields a more direct way to check the validity of \eqref{eq:Seff}, at least from the perspective of the steady-state. The steady-state(s) predicted by the leading-order DMFT dynamics $\mathcal L_\text{eff}$ admit the following purification:
\begin{align}
    |\Psi_\text{DMFT}\rangle = \sum_{m=0}^\infty \frac{1}{m!}\bigg(\frac{2\hat{K}_+}{2U\bar{n}^\text{MF} -\Delta_\text{eff}}\bigg)^m|h_0\rangle,\label{eq:psiDMFT}
\end{align}
where $\bar{n}^\text{MF}$ is any solution to the cubic self-consistency condition mentioned in the preceeding subsection. We can directly see that the exact solution indeed converges to this form as $N\to\infty$. We begin by noticing the asymptotics $(N/2U)^m \Gamma(\delta)/\Gamma(m+\delta) \underset{N\to\infty}{\sim} (2Um/N-\Delta_\text{eff})^{-m}$ for the Gamma function, which yields the asymptotic estimate
\begin{align}
    |\Psi_{\hat{T}}\rangle \underset{N\to\infty}{\sim} \sum_{m=0}^\infty \frac{1}{m!}\bigg(\frac{2\hat{K}_+}{2Um/N-\Delta_\text{eff}}\bigg)^m|h_0\rangle.
\end{align}
Notice that the above is identical to \eqref{eq:psiDMFT}, provided that we replace $m\sim 2\hat{N}_+$ with its average value. Indeed, fluctuations in $\hat{N}_+$ with respect to the exact solution $|\Psi_{\hat{T}}\rangle$ can be evaluated directly via \eqref{eq:collective} and shown to be of order $O(N^{1/2})$, so that $m/N$ becomes a deterministic variable in the large-$N$ limit. The series above thus becomes well-concentrated about $m/N\sim \bar{n}^\text{MF}$, leading to the self-consistent form \eqref{eq:psiDMFT} predicted by leading-order DMFT. 

\subsubsection*{Establishing the location of the critical point}
\begin{figure}
    \centering
    \includegraphics[width =0.95\columnwidth]{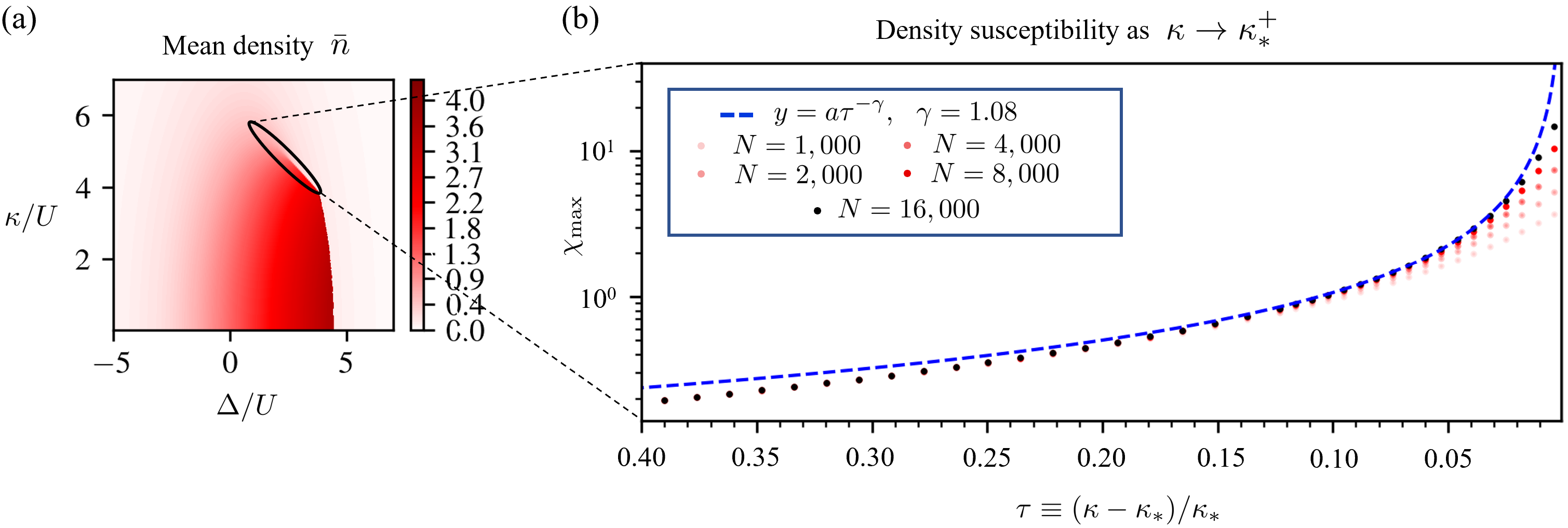}
    \caption{{\bf Confirming the location of the critical point using the exact solution}. (a)  Average density as a function of detuning $\Delta$ and loss $\kappa$, with $N=500$, $\Lambda = 0$, and $G = U$.  Phase boundaries can be seen, provided that $\kappa<\kappa_*\approx 4U$. (b) Maximum absolute value of the susceptibility as a function of $\kappa$, for $\kappa \to \kappa_*^+$. Here, $\Lambda=0$, and $G = U/10$. The polynomial fit used to estimate $\gamma$ is depicted (dashed blue line).}
    \label{fig:criticality}
\end{figure}

We will now test the claim made previously, namely that the critical point obtained from the exact solution lies exactly at the mean-field critical point, i.e. we will compute the maximum magnitude of the susceptibility 
\begin{align}
    \chi_\text{max}(\kappa,N):= \sup_{\Delta}\Bigg|\frac{\partial\bar{n}}{\partial \Delta} \Bigg|   
\end{align}
where all other parameters are held fixed. We confirm that $\chi_\text{max}$ diverges as $N\to\infty$ whenever $\kappa<\kappa_*$, and converges otherwise. Repeatedly testing for the convergence of $\chi_\text{max}$ in the thermodynamic limit for different values of $\kappa$ allows us to approximately compute the rate at which $\chi_\text{max}$ diverges, thus obtaining the critical exponent $\gamma$:
\begin{align}
    \lim_{N\to\infty}\chi_\text{max}(\kappa,N) \underset{\kappa\to \kappa_*^+}{\sim} \tau^{-\gamma},
\end{align}
with $\tau:= (\kappa-\kappa_*)/\kappa_*$. Using a very crude polynomial fitting algorithm, we estimate $\gamma \approx -1$ (see Figure \ref{fig:criticality}(b)). 

\subsection{Semiclassical limit}
We now investigate our model without the $D\equiv 0$ restriction, but in the semiclassical limit $\bar{n}\gg 1$. When the onsite photon occupation is large, the dynamics of the field amplitudes $\alpha_j(t):=\text{Tr}[\hat{\rho}(t)\hat{a}_j]$ is well captured by the semiclassical equation of motion
\begin{align}
    iu^{-1}\partial_t \beta_j &=2\lambda_j\beta^*_j +\beta_j\bigg[\bigg(2\sum_{k}|\beta_k|^2+1\bigg)-\Delta_\text{eff}/u\bigg],\label{eq:semiclassical_eq}
\end{align}
where
\begin{align}
    \beta_j^*\equiv \sum_{k=1}^N V_{j,k} \alpha_{k}^*
\end{align}
is the change-of-coordinates on the classical phase space induced by the unitary $V$ in the Autonne-Takagi factorization $M/u= V\Sigma V^T$. We now demonstrate the claim made in the main text, namely that the stable stationary states of the above dynamical system are spheres in phase space formed by the max-pairing modes. In particular, the semiclassical fixed-points satisfy the equations
\begin{align}
    0=2\lambda_j\beta^*_j +\beta_j\bigg[\bigg(2\sum_{k}|\beta_{k}|^2+1\bigg)-\Delta_\text{eff}/u\bigg]\label{eq:fixedpt_eqn}
\end{align}
From now on, for clarity, we will use the symbol $X$ to denote the set of fixed points. Note that, trivially, $\vec{0}\in X$. However, we are interested in the {\it nonzero} fixed points. Therefore, let $\vec{\beta}_\text{ss}\in X$ be a nonzero fixed point, i.e. a nonzero solution to \eqref{eq:fixedpt_eqn}. Whenever $\beta_{j,\text{ss}}\neq 0$, we can divide through by $\beta_j$ in \eqref{eq:fixedpt_eqn}, yielding a constraint on the phase $e^{i\theta_j}\equiv \beta_{j,\text{ss}}/|\beta_{j,\text{ss}}|$:
\begin{align}
    e^{-2i\theta_j} = -\frac{2R_\text{ss}^2+1 - \Delta_\text{eff}/u}{2\lambda_j} ,\label{eq:2}
\end{align}
where $R^2_\text{ss}\equiv \sum_j|\beta_{j,\text{ss}}|^2$. In particular, if $\beta_{i,\text{ss}},\beta_{j,\text{ss}}\neq 0$ is any pair of nonzero components of $\vec{\beta}_\text{ss}$, then, by taking the absolute value of the above equation,
\begin{align}
    \frac{1}{2\lambda_i} |2R^2_\text{ss}+1 - \Delta_\text{eff}/u| = \frac{1}{2\lambda_j} |2R_\text{ss}^2+1 - \Delta_\text{eff}/u|.
\end{align}
In particular, $\kappa \neq 0$ and so we can divide both sides by $|2R^2_\text{ss}+1 - \Delta_\text{eff}/u|$. Therefore, $\lambda_i=\lambda_j$. We can go even further: \eqref{eq:2} also means that $e^{2i\theta_i}=e^{2i\theta_j}$. In summary, for each $\vec{\beta}_\text{ss}\in X$, there exists a $\lambda,\theta$ such that, for all nonzero components of $\vec{\beta}_\text{ss}$,
\begin{align}
    \lambda_j=\lambda,~~~~~\beta_{j,\text{ss}}=e^{i\theta}x_j,~~~~~x_j\in \mathbb{R},
\end{align}
where $\theta$ is independent of $j$ and uniquely determined by $\lambda$ in the following way:
\begin{align}
    e^{-2i\theta} = -\frac{2R^2_\text{ss}+1 - \Delta_\text{eff}/u}{2\lambda}.\label{eq:3}
\end{align}
By taking the real and imaginary parts of the above equation, we also have
\begin{align}
2\lambda\sin 2\theta &= -\kappa/2u,\label{eq:sine}\\
2\lambda\cos 2\theta &= - (2R_\text{ss}^2+1 - \Delta/u).\label{eq:cosine}
\end{align}

\subsubsection*{Instability of solutions with $\lambda < \lambda_*$}
Let $\vec{\beta}_\text{ss}\in X$ be a nonzero fixed point, and $\lambda$ the corresponding singular value. Also, let $\lambda_*\equiv \sup_j \lambda_j$ denote the maximum singular value. We will now show that if $\lambda\neq \lambda_*$, then $\vec{\beta}_\text{ss}$ is unstable. For this purpose, it will be useful to rewrite the equations of motion in a coordinate system $(\beta_1',\dots, \beta_{N}')$ that is adapted to $\vec{\beta}_\text{ss}$, namely such that
\begin{align}
    (\beta_{1,\text{ss}}',\dots, \beta_{N,\text{ss}}') = ( e^{i\theta}R,0,\dots, 0).
\end{align}
Crucially, by the arguments in the preceeding subsection, we can achieve this via a rotation of the mode amplitudes $\beta_j$ with $\lambda_j\equiv \lambda$:
\begin{align}
    \beta_j' = \begin{cases}\sum_{\lambda_k=\lambda}A_{j,k}\beta_k&\lambda_j = \lambda\\ \beta_j&\lambda_j\neq \lambda
    \end{cases},~~~~~~~ A\in  O(s_\lambda,\mathbb{R}),
\end{align}
where $s_\lambda$ denotes the multiplicity of the singular value $\lambda$. Since the above transformation is a symmetry of the equations of motion \eqref{eq:semiclassical_eq}, the equations of motion are covariant with respect to this transformation:
\begin{align}
    iu^{-1}\partial_t \beta_j' &=2\lambda_j(\beta_j')^* +\beta_j'\bigg(2\sum_k |\beta_k'|^2+1-\Delta_\text{eff}/u\bigg).\label{eq:semiclassical_eq_newframe}
\end{align}
Now let $\delta\beta'_j\equiv \beta_j'-\beta_{j,\text{ss}}'$ denote the fluctuations about $\vec{\beta}_\text{ss}$. Assuming these fluctuations are small, we can obtain linearized equations of motion for these fluctuations:
\begin{align}
     iu^{-1}\partial_t \delta \beta_j'&=2\lambda_j\bigg((\delta\beta_j')^*- \frac{\lambda}{\lambda_j}e^{-2i\theta}\delta \beta_j'\bigg)+ 2\beta_{j,\text{ss}}'\sum_k ((\beta_{k,\text{ss}}')^*\delta \beta_k' +c.c.) + O(\delta\beta'^2)
\end{align}
where we have implicitly used \eqref{eq:2}. We now wish to argue that, if $\lambda\neq \lambda_*$, then the Hurwitz criterion fails, that is, the associated dynamical matrix contains an eigenvalue with positive real part. It suffices to examine the stability of the fluctuations $\delta\beta_j'$ for $j\neq 1$, which evolve within this linear approximation as follows:
\begin{align}
    iu^{-1}\partial_t  \delta\beta_j'&=2\lambda_j\bigg((\delta \beta_j')^*- \frac{\lambda}{\lambda_j}e^{-2i\theta}\delta \beta_j'\bigg)\\
    &=2\lambda_j (\delta\beta_j')^* -\frac{\kappa}{2}\delta\beta_j'-i\bigg(\Delta-u(2R^2_\text{ss}+1)\bigg)\delta\beta_j',
\end{align}
where we have shown explicitly that this is the equation of motion for a detuned parametric amplifier, with detuning modified by the presence of the Hubbard interaction $u$. Proceeding with the calculation, we then split the fluctuations into real and imaginary parts via $e^{-i\theta}\delta\beta_j' = \delta x_j + i\delta y_j$, and obtain the equations of motion
\begin{align}
    u^{-1}\partial_t(\delta x_j + i\delta y_j) = -2\lambda_j e^{-2i\theta} \bigg((1+\lambda/\lambda_j)\delta y_j +i(1-\lambda/\lambda_j) \delta x_j\bigg).
\end{align}
The corresponding eigenvalues of the dynamical matrix, using \eqref{eq:sine}, are
\begin{align}
    \gamma_j^\pm = -\kappa/2\pm \sqrt{(\kappa/2)^2- 4u^2(\lambda^2-\lambda_j^2)},\label{eq:previous_analysis}
\end{align}
Therefore, if $\lambda_j > \lambda$, then the eigenvalue $\gamma_j^+$ has positive real part. Therefore, if $\lambda\neq \lambda_*$, then $\vec{\beta}_\text{ss}$ is unstable. We also have a partial converse statement: if $\lambda_j <\lambda$, then the fluctuations $\delta \beta_j'$ are stable.

\subsubsection*{Stability of solutions with $\lambda =\lambda_*$}
Let $\vec{\beta}_\text{ss}\in X$ be a nonzero fixed point with corresponding singular value $\lambda_*$. We will now investigate the conditions under which $\vec{\beta}_\text{ss}$ is stable. By \eqref{eq:previous_analysis}, the fluctuations $\delta \beta_j'$ for $j=2,3,\dots N$ are all appropriately damped. We thus must investigate the stability of the remaining fluctuations $\delta\beta_1'$, which have the following linearized equations of motion:
\begin{align}
    u^{-1}\partial_t (e^{-i\theta} \delta\beta_1') =-4\lambda_*e^{-2i\theta}\delta y_1 - 4iR_\text{ss}^2\delta x_1.
\end{align}
The corresponding eigenvalues of the dynamical matrix, using (\ref{eq:sine}-\ref{eq:cosine}), are
\begin{align}
    \gamma^\pm_j = -(\kappa/2)\pm \sqrt{(\kappa/2)^2 - 8u^2R_\text{ss}^2(2R_\text{ss}^2+1)+8u\Delta R_\text{ss}^2}. 
\end{align}
Therefore, $\vec{\beta}_\text{ss}$ is stable if and only if 
\begin{align}
    8u^2R_\text{ss}^2(2R_\text{ss}^2+1)+8u\Delta R_\text{ss}^2> 0, \label{eq:criterion}
\end{align}
which happens if and only if $u^2(2R_\text{ss}^2+1)-u\Delta > 0$. To see whether this criterion is satisfied, we must use the fact that $\vec{\beta}_\text{ss}$ is a fixed point in order to obtain an additional constraint on $R_\text{ss}$. In particular, taking the absolute value squared of \eqref{eq:2} yields a quadratic equation for $R_\text{ss}^2 + 1$, with two possible solutions:
\begin{align}
    2R_\text{ss}^2+1 = \Delta/u \pm \sqrt{(2\lambda_*)^2-(\kappa/2)^2}.
\end{align}
Therefore, criterion \eqref{eq:criterion} is satisfied if and only if
\begin{align}
    R_\text{ss} =  \sqrt{\frac{\Delta/u -1 + \sqrt{(2\lambda_*)^2-(\kappa/2)^2}}{2}}>0.\label{eq:theradius}
\end{align}
Finally, let $\beta_j$ for $j=1,2,\dots, s$ denote the eigenmodes corresponding to the maximum singular value $\lambda_*$, i.e. the so-called "max-pairing modes". Since rotations $A\in O(s,\mathbb{R})$ of the max-pairing modes are symmetries of the equations of motion, any such rotation $A$ must send a stable fixed point to another stable fixed point. Therefore, when the inequality \eqref{eq:theradius} is satisfied, the space $X^\text{stab.}\subset X$ of nonzero stable fixed points is a sphere:
\begin{align}
    X^\text{stab.} = e^{i\theta}\bigg\{x_j\in \mathbb{R}^s:~~~\sum_{\lambda_j = \lambda_*}x_j^2 = R_\text{ss}^2\bigg\}
\end{align}
In particular, when $s>1$, the fluctuations tangent to $X^\text{stab.}$ are Goldstone modes, i.e. zero-modes for the linearized dynamics. We can verify this explicitly by fixing a point $\vec{\beta}_\text{ss}\in X^\text{stab.}$, and expanding the linearized equations of motion for the resulting fluctuations $e^{-i\theta}\delta \beta_j' = \delta x_j + i\delta y_j$, in the coordinate frame adapted to $\vec{\beta}_\text{ss}$:
\begin{align}
    iu^{-1}\partial_t(\delta x_j + i\delta y_j)= -4i\lambda_*e^{-2i\theta}\delta y_j,~~~~j=2,\dots, s
\end{align}
In particular, the real-components $\delta x_j \in T_{\vec{\beta}_\text{ss}}X^\text{stab.}$ of the fluctuations are zero modes of the dynamical matrix, as was expected.

\newpage

\section{Mathematical background}\label{sec:the_sec}
\subsection{Proof of the $SU(1,1)$ decomposition theorem}
We now establish the decomposition \eqref{eq:the_decomp}. To make our task easier, we establish the desired decomposition for a dense subspace of the Hilbert space. The corresponding decomposition for the full Hilbert space can then be proven using standard functional-analytic techniques. \\

In particular, let $V^{(j)}$ be the local nonunitary $SU(1,1)$ representations defined in \eqref{eq:localreps}. We define $W^{(j)}\subset V^{(j)}$ to be the {\it algebraic} part of each representation, that is, the part of the representation consisting of finite linear superpositions of Fock staes:
\begin{align}
    W^{(j)} := \bigg\{\sum_n a_n \hat{\beta}_{j,+}^{\dagger n}|h_0\rangle\in V^{(j)},~~~~~~\text{finitely many~}a_n~\text{nonzero}\bigg\}
\end{align}
The above subspaces allow us to conveniently establish the following theorem:\\

\noindent {\bf Theorem}. {\it The global nonunitary $SU(1,1)$ representation $\otimes_j W^{(j)}$ decomposes into irreducible subrepresentations as follows}:
\begin{align}
    \otimes_j W^{(j)}  \simeq \bigoplus_{l=0}^\infty\Bigg(\bigoplus_{p=1}^{d_l} W_l^{(p)}\Bigg), \label{eq:the_decomp1}
\end{align}
{\it where the spaces $W^{(p)}_l\subset V^{(p)}_l$ are defined analogously}:
\begin{align}
    W^{(p)}_l := \bigg\{\sum_n a_n \hat{K}_+^n|h_l^{(p)}\rangle,~~~~~~\text{finitely many~}a_n~\text{nonzero}\bigg\}.
\end{align}

\noindent {\it Proof}. This can be proved by going to the Segal-Bargmann representation. Within this representation, a finite-boson number state is represented as a polynomial:
\begin{align}
    \sum_n a_n \hat{\beta}_{j,+}^{\dagger n}|h_0\rangle \in W^{(j)}\to \sum_n a_nx_j^n\in \mathbb{C}[x_j],
\end{align}
with $\mathbb{C}[x_j]$ the univariate polynomial ring generated by $x_j$. It then follows that the tensor product $\otimes_jW^{(j)}$ is the multivariate polynomial ring $\mathbb{C}[x_1,\dots, x_N]$ generated by the indeterminates $x_1,\dots x_N$. Finally, creation and annihilation operators are represented by partial differential operators:
\begin{align}
    \hat{\beta}_{j,+}\to \frac{\partial}{\partial x_j},~~~~~~\hat{\beta}_{j,+}^\dagger\to x_j.
\end{align}
We first establish our decomposition under the assumption that the singular values $\lambda_j \equiv \lambda$ are all completely degenerate, and then generalize the argument to the generic non-degenerate case $\lambda_i\neq\lambda_j$. In the degenerate case, the global $SU(1,1)$ representation takes the simple form
\begin{align}
    \hat{K}_+ = \frac{\lambda}{2}
    \sum_j x_j^2,~~~~\hat{K}_- = \frac{1}{2\lambda} \sum_j \frac{\partial^2}{\partial x_j^2}\label{eq:therep}
\end{align}
Therefore, within the Segal-Bargmann representation, our decomposition theorem is equivalent to the following decomposition of the polynomial ring $\otimes_jW^{(j)}$:
\begin{align}
    \mathbb{C}[x_1,...,x_N]  \simeq \bigoplus_{l=0}^\infty\Bigg(\bigoplus_{p=1}^{d_l} \mathbb{C}[R^2]h_l^{(p)}\Bigg), \label{eq:the_decomp2}
\end{align}
where $R^2 = \sum_j x_j^2$, and $h_l^{(1)},...,h_l^{(d_l)}$ is some orthonormal basis of the space of harmonic homogeneous polynomials of degree $l$. When interpreted pointwise, the above isomorphism reads as a harmonic expansion of a fixed polynomial:
\begin{align}
    p(x_1,...,x_N) = \sum_{l=0}^\infty \sum_{p=1}^{d_l} q^{(p)}_{l}(R^2) h_{l}^{(p)}(\vec{x}) \label{eq:harmdecomp_pointwise}
\end{align}
where the $q^{(p)}_l$ are univariate polynomials. The decomposition \eqref{eq:the_decomp2} is then equivalent to the statement that the above mapping, interpreted as a mapping from the RHS to the LHS, is an isomorphism of $SU(1,1)$-representations. The representation \eqref{eq:therep} acts the same on both sides, so it suffices to establish the isomorphism at the vector space level, i.e. prove that the above mapping is both injective and surjective. For injectivity, it suffices to show that the subspaces $\mathbb{C}[R^2]h_{l}^{(p)}$ are all mutually nonintersecting (except at $\{0\}$), which can be verified by direct calculation:
\begin{align}
    \langle R^{2n}h_l^{(p)}|R^{2m}h_{l'}^{(q)}\rangle = \bigg(\frac{2}{\lambda}\bigg)^{4n}\delta_{p,q}\delta_{l, l'}\delta_{n,m} n!(N/2+l)_n\label{eq:superselection_2}
\end{align}
To demonstrate surjectivity, we must demonstrate that, for {\it any} polynomial $p$ on the LHS of \eqref{eq:harmdecomp_pointwise}, a decomposition of the form given on the RHS exists. This seems considerably more challenging to establish. However, here we are helped by a basic fact from harmonic analysis: \\

\noindent {\bf Lemma}. {\it let $p$ denote a homogeneous polynomial of degree $l$. Then}
\begin{align}
    p(x_1,...,x_N) = h(\vec{x}) + R^2 q(x_1,...,x_N),
\end{align}
{\it where $q$ is a homogeneous polynomial of degree $l-2$, and $h$ is a homogeneous harmonic polynomial of degree $l$.}\\

\noindent A proof is usually given in standard textbooks on harmonic function theory (see, e.g. \cite{axlersheldonHarmonicFunctionTheory2001}). In any case, iterating the above lemma, we obtain, for any homogeneous polynomial $p$ a decomposition
\begin{align}
    p(x_1,...,x_N) = \sum_{m=0,1,...} R^{2m} h_{\deg p -2m}(\vec{x}), \label{eq:protodecomp}
\end{align}
where $h_l$ denotes a homogeneous harmonic polynomial of degree $l$, so that the expansion \eqref{eq:harmdecomp_pointwise} can be established simply by writing out the polynomial on the LHS of \eqref{eq:harmdecomp_pointwise} as a sum of homogeneous components, and then expanding the harmonic polynomials appearing on the RHS of \eqref{eq:protodecomp} into a basis.\\

The preceeding arguments constitute a proof of the theorem for the degenerate case $\lambda\equiv \lambda_j$. Therefore, all that is left is to reproduce the above proof in the non-degenerate case $\lambda_i\neq \lambda_j$. Luckily, the proof for the non-degenerate case follows immediately from the degenerate case. In particular, we can write the $SU(1,1)$ representation as
\begin{align}
    \hat{K}_+ = \frac{1}{2}
    \sum_j y_j^2,~~~~\hat{K}_- = \frac{1}{2} \sum_j \frac{\partial^2}{\partial y_j^2}\label{eq:therep_nonunitary}
\end{align}
where we have made the change of variables $x_j \to y_j:= \lambda_j^{1/2}x_j$. In particular, just by making the replacements $x_j\to y_j$ in \eqref{eq:the_decomp2}, we obtain the desired result:
\begin{align}
    \mathbb{C}[y_1,...,y_N]  \simeq \bigoplus_{l=0}^\infty\Bigg(\bigoplus_{p=1}^{d_l} \mathbb{C}[R^2]h_l^{(p)}\Bigg), \label{eq:the_decomp3}
\end{align}
with $R^2$ and $h_l^{(p)}$ defined just as in \eqref{eq:the_decomp2}, but with the replacements $x_j\to y_j$. This completes the proof of our theorem in the non-degenerate case.


\subsection{Exact solution for the steady-state Wigner function}
We now compute a closed form for the Wigner function $W_\text{ss}$ of the steady-state density matrix $\hat{\rho}_\text{ss}$. The arguments in \cite{roberts_driven_2019} generalize in a straightforward manner to the $N$-mode case. In particular, the steady-state Wigner function for our system satisfies the identity
\begin{align}
        W_\text{ss}(\vec{\alpha})& = 2^NQ_{|\Psi_+\rangle}\bigg(\sqrt{2}\vec{\alpha}\bigg),\label{eq:Qfunction}
\end{align}
where $Q_{|\Psi_+\rangle}$ is the Husimi-Q representation of the $N$-mode pure state $|\Psi_+\rangle$. Therefore, to express the Wigner function $W_\text{ss}$ of the steady state in closed form it suffices to express the Husimi-Q representation of $|\Psi_+\rangle$ in closed form. This task is solved easily via the Segal-Bargmann representation \cite{bargmann_hilbert_1961,bargmann_hilbert_1967} of $|\Psi_+\rangle$, which can be calculated in a straightforward manner from \eqref{eq:ansatz}:
\begin{align}
    \Psi_\text{SB}(\vec{\alpha}) =\frac{\,_0F_1\big(\delta; -\frac{1}{2}\sum_{ij}\frac{M_{ij}}{u}\alpha_i\alpha_j\big)}{\sqrt{\mathcal N}},~~~~~\mathcal N=\sum_{l=0}^\infty \frac{\Phi_l(\vec{\lambda};N)}{(\delta)_l(\delta^*)_l}.
\end{align}
Now, the $Q$-function of a pure state $|\Psi\rangle$ is given by $Q_{|\Psi\rangle}(\vec{\alpha}) =\pi^{-N}|\Psi_\text{SB}(\vec{\alpha}^*)|^2 e^{-|\vec{\alpha}|^2}$, where $\Psi_\text{SB}$ is the Segal-Bargmann representation of $|\Psi\rangle$. Therefore, \eqref{eq:Qfunction} yields
\begin{align}
    W_\text{ss}(\vec{\alpha})&=\frac{1}{\mathcal N}\bigg(\frac{2}{\pi}\bigg)^N\bigg|\,_0F_1\bigg(\delta ;-\sum_{ij}\frac{M_{ij}}{u}\alpha_i^*\alpha_j^* \bigg)\bigg|^2e^{-2|\vec{\alpha}|^2},\label{eq:wigner}
\end{align}

\begin{figure}
    \centering
    \includegraphics[width = 1.0\columnwidth]{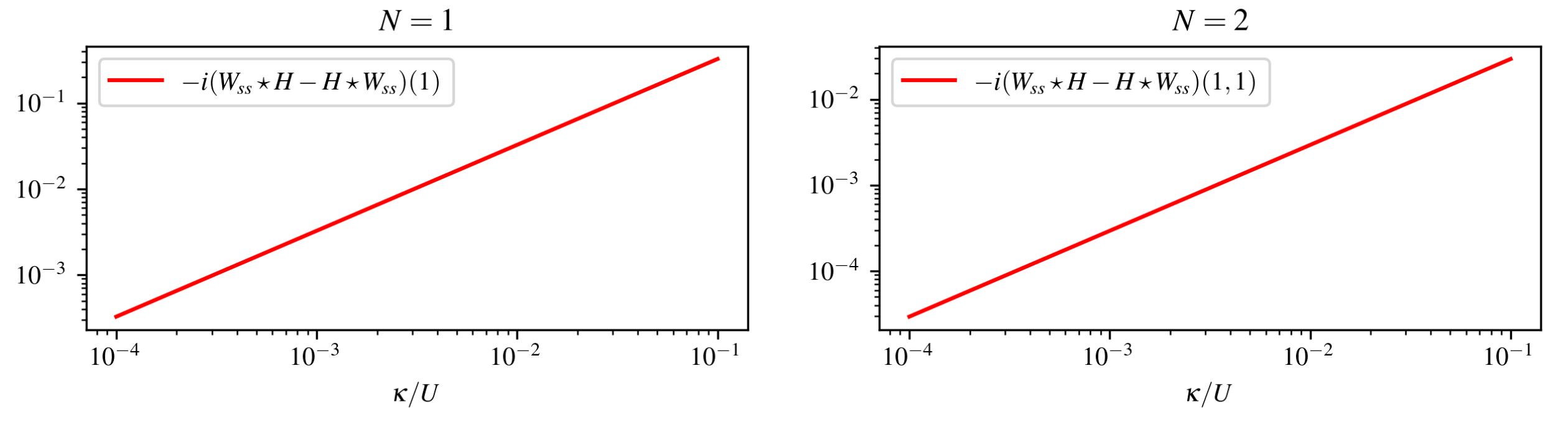}
    \caption{{\bf Nonthermal nature of the steady state}. The exact solution \label{eq:wigner_unitary} for the Wigner function can be used to symbolically check that the Hamiltonian and steady state do not commute. Here, we evaluate the phase-space commutator $H\star W_\text{ss}- W_\text{ss}\star H$ in the case $G=\kappa = U,$ and $\Lambda =0$, for the cases (a) $N=1$, in which case we choose to evaluate the result at the phase space point $\alpha = 1$ and (b) $N=2$, in which case we choose to evaluate the result at the phase space point $\alpha_1=\alpha_2 = 1$. Note that, since $H,W_\text{ss}$ are both purely real, the result is always purely imaginary. As expected, the result always vanishes in the limit $\kappa\to 0^+$.}
    \label{fig:moyal}
\end{figure}

\subsubsection*{Scaling limit for the Wigner function}
We now investigate the expression for the Wigner function in the high-density limit $M_{ij}/u\to \infty$. Note that the expression \eqref{eq:wigner} for the Wigner function is non-negative, and thus can be interpreted as a bona-fide probability measure. In the limit $M_{ij}/u\to \infty$, this probability measure will become supported on larger and larger regions of phase space, and so it is useful to re-scale the phase space in such a way that the resulting rescaled distribution converges to a limit. \\

The correct scaling turns out to be $\beta_j := \sqrt{-\lambda_*}\tilde{\beta}_j$, where $\lambda_*\equiv \sup_j\lambda_j$ is the maximum singular value appearing in the Autonne-Takagi factorization $M/u=V\Sigma V^T$, and 
\begin{align}
    \beta_j^*\equiv \sum_{k=1}^N  V_{j,k}\alpha_k^*   
\end{align}
is the change-of-coordinates on the classical phase space of our system, induced by the unitary $V$ appearing in the factorization. The explicit form of $V$ can be recovered from the spectral decomposition of the Laplacian of our underlying connectivity graph. One can show that, at least when $\Delta = 0$, $\kappa=0^+$, in which case the Bessel function sitting inside the absolute value in \eqref{eq:wigner} becomes a hyperbolic cosine, the steady-state Wigner function $W_\text{ss}(\tilde{\beta})$ limits to a uniform distribution on the sphere
\begin{align}
    S\equiv \bigg\{(\tilde{\beta}_1,\dots, \tilde{\beta}_s,0,\dots, 0)\in\mathbb{R}^N:~~\sum_{j=1}^s\tilde{\beta}_j^2=1\bigg\},
\end{align}
where $\beta_1,\dots ,\beta_s$ are the max-pairing modes. This can be established by expanding the Wigner function \eqref{eq:wigner} into a sum of four exponentials, and then solving the associated saddle-point equations.

\subsection{Using the Wigner function to verify the nonthermal character of the steady state}
{The exact solution for the Wigner function can be used to explicitly showcase the nonequilibrium character of the steady state. One non-thermal feature of the steady state is that it does not commute with the Hamiltonian, that is, $[H, \rho_\text{ss}]\neq 0$. Therefore, the steady state cannot be written as $\exp(-\beta \hat{H})$ for some $\beta$. This can be verified efficiently when $\Lambda=0$, in which case we can write down the closed-form solution
\begin{align}
    W_\text{ss}(\vec{\alpha}) &=\bigg(\frac{2}{\pi}\bigg)^N\frac{\big|\,_0F_1\big(\delta ;-2\lambda\sum_j\alpha_j^{*2} \big)\big|^2}{\,_1F_2(N/2;\delta,\delta^*;\lambda^2)}e^{-2|\vec{\alpha}|^2},\label{eq:wigner_unitary}
\end{align}
with $\lambda =NG/U$ in this case corresponding to the unique singular value of the pairing matrix. To show that the steady state and the Hamiltonian do not commute, we pass to the phase-space formulation of quantum mechanics. In the phase space formulation of quantum mechanics, the noncommutativity of the steady state and the Hamiltonian is equivalent to the statement that
\begin{align}
    W_\text{ss}\star H - H\star W_\text{ss}\neq 0, \label{eq:commutator}
\end{align}
where $\star$ is the Moyal product \cite{moyal_1949}, and $H$ denotes the symmetrically-ordered (i.e. Weyl) symbol of the Hamiltonian. Because $H$ is a polynomial, the following derivative expansion terminates at a finite order and hence can be calculated symbolically in closed form using a simple computer algebra program:
\begin{align}
    (f\star g)(\vec{\alpha}) = \exp\bigg(-\frac{1}{2}\sum_{j=1}^N \bigg(\frac{\partial}{\partial x_j^*}\frac{\partial}{\partial y_j} - \frac{\partial}{\partial x_j}\frac{\partial}{\partial y_j^*}\bigg)\bigg)f(\vec{x})g(\vec{y})\bigg|_{\vec{x}=\vec{y}=\vec{\alpha}}
\end{align}
Since both the Wigner function and Hamiltonian are generically smooth functions of $\vec{\alpha}$, the phase-space function  \ref{eq:commutator} is generically a smooth function of $\vec{\alpha}$. In Fig. \ref{fig:moyal} we exhibit a single point where this phase-space function is non-vanishing.}

\newpage

\section{Experimental realization using superconducting circuits ($D=0$)}

\begin{figure}[h]
    \centering
    \includegraphics[width = 0.22\columnwidth]{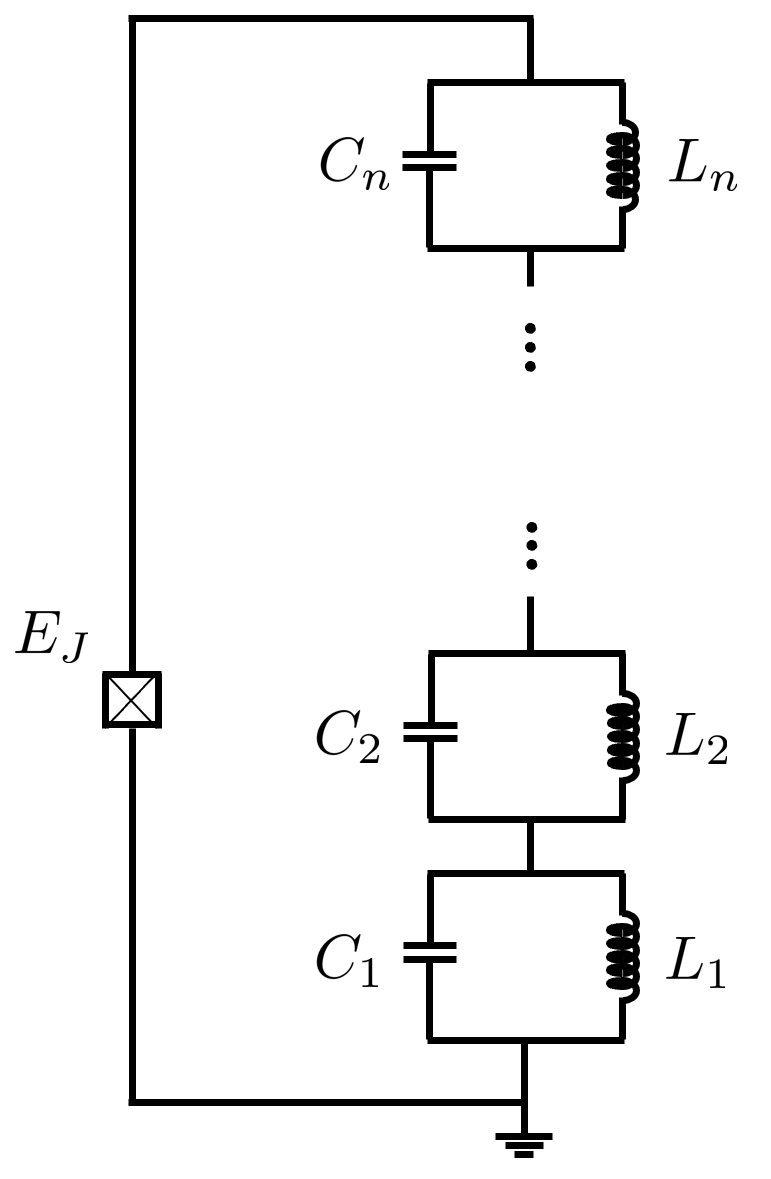}
    \caption{{\bf  First iteration of the circuit (no driving)}. A Josephson tunnel junction is placed in parallel with a chain of LC oscillators to provide a global Hubbard interaction.}
    \label{fig:circuitFig1}
\end{figure}
Our $D=0$ model is relatively easily realizable using a simple superconducting circuit with only three nonlinear elements. To see this, we first attempt to realize the $D = 0$ Hamiltonian without the driving. We begin by placing a chain of $N$ uncoupled LC oscillators in series. Via examination of Kirchoff's current laws, the Hamiltonian that describes the equations of motion for the chain is the following:
\begin{align}
    H = \sum_j \bigg(\frac{Q_j^2}{2C_j} + \frac{\Phi_j^2}{2L_j}\bigg),
\end{align}
where here, $Q_j$ denotes the charge stored on the $j$th capacitor, and $\Phi_j$ denotes the time-integrated voltage across each inductor. One may diagonalize this Hamiltonian by defining dimensionless creation- and annihilation operators 
\begin{align}
    \hat{\Phi}_j := \Phi_j^\text{zpf}(\hat{a}_j+\hat{a}_j^\dagger),~~~~~~\hat{Q}_j := -iQ_j^\text{zpf} (\hat{a}_j-\hat{a}_j^\dagger),
\end{align}
where here, $Q_j^\text{zpf}, \Phi_j^\text{zpf}$ are the zero-point vacuum fluctuations of the charge and phase across the inductive and capacitive branches of each LC oscillator.

To add an infinite-range Bose-Hubbard interaction, we place a Josephson junction in parallel with the chain of oscillators (c.f. Figure \ref{fig:circuitFig1}). The Josephson junction, in the limit of extremely weak junction capacitance $C_J\ll C_j$, can be modelled accurately to leading order via the following interaction Hamiltonian:
\begin{align}
    \hat{H}_\text{int} = E_J\cos\bigg(\sum_j \varphi_j (\hat{a}_j+\hat{a}_j^\dagger)\bigg),
\end{align}
where $\varphi_j := \Phi_j^\text{zpf}/2\pi\Phi_0$. We now tune the dimensionless phase fluctuations $\varphi_j$ to be parametrically small and uniform across all modes, that is $\varphi_j\equiv \varphi \ll 1$. Note that this can be done without constraining the resonant frequencies of each oscillator. Taylor's theorem then says that
\begin{align}
    \hat{H}_\text{int} = -\frac{E_J\varphi^2}{\hbar 2!}\bigg(\sum_j \hat{a}_j + h.c.\bigg)^2 + \frac{E_J\varphi^4}{\hbar 4!}\bigg(\sum_j\hat{a}_j + h.c.\bigg)^4 + O(\varphi^6).
\end{align}
We then go into a rotating frame with respect to the free Hamiltonian $H_\text{free}:= \hbar \sum_j \omega_j \hat{n}_j$, where $\omega_j$ is the bare resonance frequency of each LC resonator in the chain. Now we choose the resonant frequencies of each mode so that the fundamental frequency $\Omega$ of the rotating-frame Hamiltonian is much larger than the rate $\hbar^{-1}E_J\varphi^2$. In this regime, the rotating-wave approximation is valid and yields
\begin{align}
    \hat{H}_\text{RWA} = -E_J\varphi^2\sum_j \hat{n}_j+\frac{E_J\varphi^4}{2}\bigg(\sum_j\hat{n}_j\bigg)^2.
\end{align}

\subsection{Adding a two-photon drive}

\begin{figure}
    \centering
    \includegraphics[width = 0.35\columnwidth]{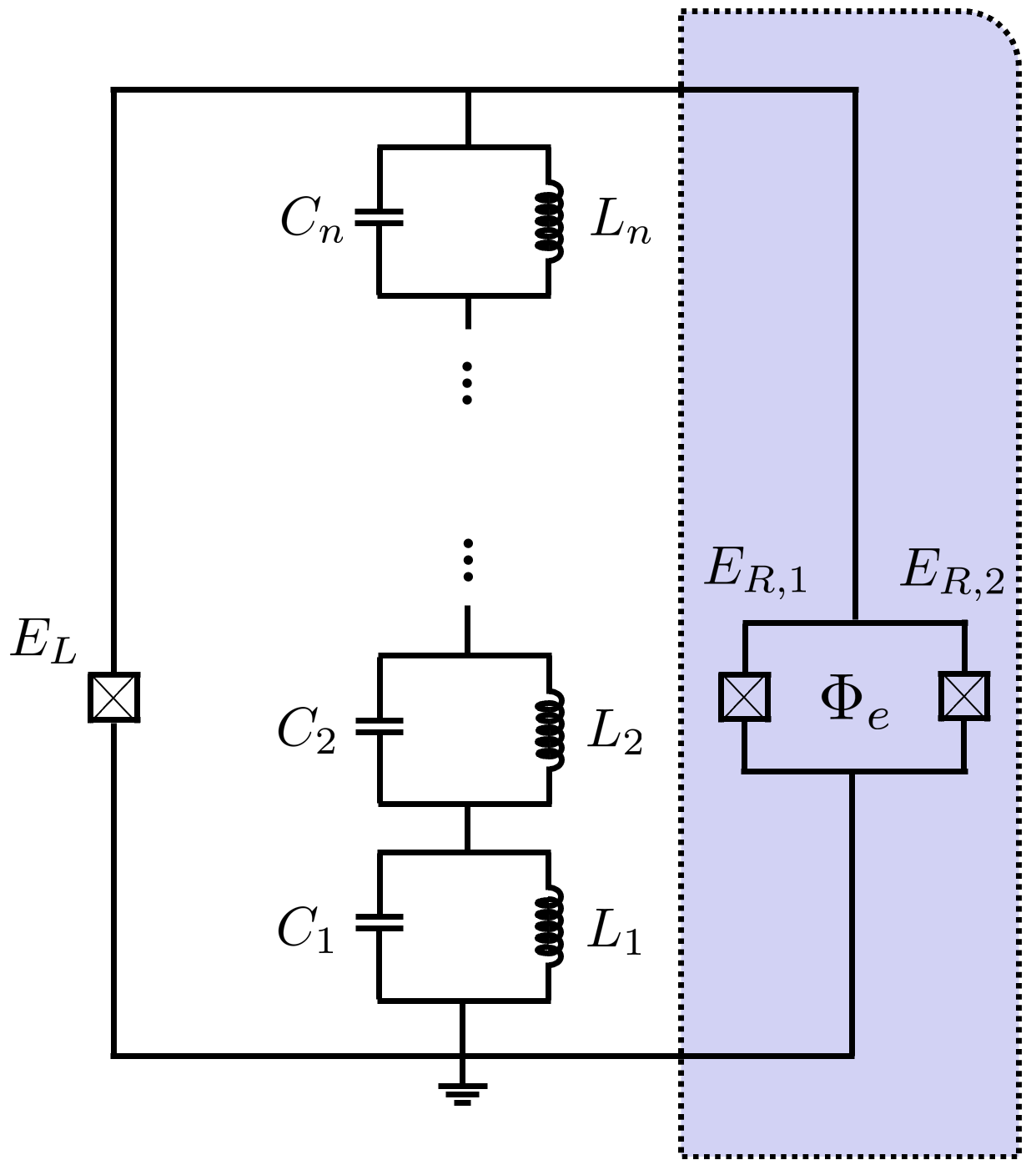}
    \caption{{\bf Circuit incorporating coherent two-photon driving}.  The global parametric drive is supplied by a flux-tunable transmon (blue shaded region).}
    \label{fig:my_label}
\end{figure}

To obtain the $D=0$ Hamiltonian for our model, we just have to add two-photon driving to the above scheme. To do this, we play the same trick as above: this time we add a symmetric SQUID in parallel with the oscillator chain. Assuming that the junction capacitances in the SQUID are also much smaller than the capacitances present in the oscillator chain, the new interaction Hamiltonian is simply 
\begin{align}
    \hat{H}_\text{int} = E_L \cos \bigg(\varphi\sum_j \hat{a}_j +h.c.\bigg) + 2E_R \cos \bigg(\frac{\Phi_e}{2\pi \Phi_0}\bigg)\cos \bigg(\varphi\sum_j \hat{a}_j +h.c.\bigg),
\end{align}
where $E_R = E_{R,1},E_{R,2}$ is the Josephson energy of each junction in the symmetric SQUID. We then choose to drive the SQUID in such a way that
\begin{align}
    \frac{\Phi_e}{2\pi\Phi_0} = \frac{\pi}{2}-\epsilon_p(t).
\end{align}
We also assume that $E_R \ll E_L$. As a result, we can truncate the expansion of the SQUID potential at quadratic order, while continuing to truncate the expansion of the left junction at quartic order:
\begin{align}
    \hat{H}_\text{int} = -E_R\varphi^2 \epsilon_p(t) \bigg(\sum_j \hat{a}_j + h.c.\bigg)^2 - \frac{E_L\varphi^2}{2!}\bigg(\sum_j\hat{a}_j + h.c.\bigg)^2 + \frac{E_L\varphi^4}{4!}\bigg(\sum_j \hat{a}_j + h.c.\bigg)^4+  O(E_L\varphi^6) + O(E_R\varphi^4).
\end{align}
By modulating the pump amplitude appropriately via 
\begin{align}
    \epsilon_p(t) := \epsilon_0 \sum_j \cos (2\omega_j -2\omega_p) t,
\end{align}
and going into a rotating frame with respect to the free Hamiltonian $\hat{H}_\text{free} = \hbar \sum_j (\omega_j - \omega_p) \hat{n}_j$, and again assuming that the mode frequencies are all appropriately detuned from each other, we obtain the following rotating-wave Hamiltonian:
\begin{align}
    \hat{H}_\text{RWA} = \frac{E_L\varphi^4}{2}\bigg(\sum_j\hat{n}_j\bigg)^2 + \sum_j (\hbar \omega_p - E_L\varphi^2)\hat{n}_j - E_R\varphi^2 \epsilon_0 \sum_j (\hat{a}_j^2 +h.c.)
\end{align}
We thus obtain the exact parameters of our solvable model in the regime $D=0$ (in SI units!):
\begin{align}
    U &= \frac{NE_L\varphi^4}{2\hbar},~~~~G = -\frac{E_R\varphi^2\epsilon_0}{\hbar},~~~~\Delta = \omega_p-\frac{E_L\varphi^2}{\hbar},~~~~\varphi = \sqrt{\frac{\hbar}{2}}\frac{(L_j/C_j)^{1/4}}{2\pi \Phi_0} \equiv \text{const.} 
\end{align}

\subsection{Effect of junction capacitances}
We now compute the corrections to the Hamiltonian due to junction capacitances, and demonstrate that these capacitances can be neglected when they are much smaller than the capacitances in the oscillator chain. To simplify the analysis we assume the junction capacitances in the symmetric SQUID are the same, and define a new parameter $C_\Sigma := C_L+2C_R$ corresponding to the sum of the three junction capacitances. The Maxwell capacitance relation of the circuit, assuming the junction capacitances are all zero, is
\begin{align}
    \vec{q} = C_0 \dot{\vec{\phi}},~~~~~~~~ C_0:= \begin{pmatrix}C_1 &-C_2&0&\cdots & 0\\ -C_2&C_2+C_3 &-C_3&&\vdots\\ 0&-C_3&C_3+C_4 &-C_4&\\ \vdots &&-C_4&\ddots&\\ &&&&-C_{N-1}\\ 0&\cdots &&-C_{N-1}&C_N\end{pmatrix},
\end{align}
where the nodal charges $q_j$ can be expressed in terms of the charges $Q_j$ in the capacitance chain via 
\begin{align}
    Q_{N-j} = \sum_{k=0}^jq_{N-k},
\end{align}
With the junction capacitances included, the new capacitance matrix is obtained in a very simple manner from $C_0$:
\begin{align}
    C = C_0 + \begin{pmatrix}0&\cdots &0\\ \vdots&&\vdots \\ 0&\cdots& C_\Sigma\end{pmatrix}.
\end{align}
To obtain the corrections to the Hamiltonian that were neglected in the previous analysis, we use the Sherman-Morrison formula, which is {\it exact}:
\begin{align}
    C_{k,k}^{-1} = (C_0^{-1})_{k,k'}+\frac{C_\Sigma(C_0^{-1})_{k,N}(C_0^{-1})_{N,k'}}{1+C_\Sigma (C_0^{-1})_{N,N}},
\end{align}
so that the correction to the Hamiltonian is rigorously
\begin{align}
    \delta \hat{H} = \frac{C_\Sigma}{2}\sum_{k,k'}\frac{(C_0^{-1})_{k,N}(C_0^{-1})_{N,k'}}{1+C_\Sigma (C_0^{-1})_{N,N}}q_k q_{k'},
\end{align}
which goes to zero as $C_\Sigma(C_0^{-1})_{i,j} \to 0$,  as expected.

\newpage

\bibliography{supplemental}